\documentclass[11pt]{article}
\usepackage{graphicx} % Use pdf, png, jpg, or eps§ with pdflatex; use eps in DVI mode
\usepackage{color}
\usepackage{xcolor}
\usepackage{hyperref}
\usepackage{bookmark}
\bookmarksetup{depth=2}
\usepackage{amssymb}
\usepackage{amsmath}
\usepackage{mathtools} % more complete amsymb
\usepackage{float}
\usepackage{pgfplotstable} % build table from file content
\newlength\Myfigwd
\usepackage{cleveref}
\crefname{figure}{}{figures}
\creflabelformat{figure}{#2\txtkublue{\textsf{\textbf{\footnotesize Figure #1}}}#3}
\crefname{table}{}{tables}
\creflabelformat{table}{#2\txtkublue{\textsf{\textbf{\footnotesize Table #1}}}#3}
\crefname{section}{}{sections}
\creflabelformat{table}{#2\txtkublue{\textsf{\textbf{\footnotesize Section #1}}}#3}
\usepackage{booktabs}
\usepackage{array}
\usepackage{tocloft} % Provides control over the typography of the Table of Contents, List of Figures and List of Tables, and the ability to create new ‘List of ...’. The ToC \parskip may be changed.
\usepackage{colortbl}
\usepackage[stable]{footmisc}   % allow footnotes in section headings
\usepackage[normalem]{ulem}               % struck-out in mathmode
\usepackage{braket}
\usepackage{url}
\usepackage{upgreek}
\usepackage{bm}
\usepackage{fullpage}
\usepackage{cite}
\usepackage{subcaption}
\usepackage{outlines}
\usepackage{appendix}
\usepackage{chngcntr}
\usepackage{etoolbox}
\usepackage{xspace}
\usepackage{multirow}
\usepackage{authblk}            % affiliation
\usepackage{tikz}
\usepackage{units}              % \nicefrac{}{}
\usepackage{ifthen}             % allow conditionals in newcommands
\usepackage{pdfpages}           % add entire pdfpages

\numberwithin{equation}{section}
\numberwithin{figure}{section}

\newcommand\ignore[1]{} % ignore content

\newcommand{\m}[1]{\ensuremath{#1}}

\newcommand\n{\nonumber}

\newcommand{\that}{{\hat t}}
\newcommand{\shat}{{\hat s}}
\newcommand{\uhat}{{\hat u}}
\newcommand{\tlt}{\bar{t}}
\newcommand{\tls}{\bar{s}}
\newcommand{\tlu}{\bar{u}}

\newcommand{\be}{\begin{equation}}
\newcommand{\ee}{\end{equation}}
\newcommand{\bea}{\begin{eqnarray}}
\newcommand{\eea}{\end{eqnarray}}
\newcommand{\eq}[1]{\label{eq:#1}}
\newcommand{\fig}[1]{\label{fig:#1}}
\newcommand{\reffig}[1]{Fig.~\ref{fig:#1}}
\renewcommand{\refeq}[1]{Eq.~\ref{eq:#1}}
\newcommand{\intd}{\int\!\!\mathrm{d}} % integral mesure symbol
\newcommand{\dd}{\,\mathrm{d}} % integral mesure symbol
\renewcommand{\vec}[1]{\ifthenelse{\equal{#1}{\ell}}{\boldsymbol{#1}}{\mathbf{#1}}}
\newcommand{\abs}[1]{\left\lvert #1\right\rvert}
\newcommand{\lr}[1]{{\left({#1}\right)}}
\renewcommand{\epsilon}{\varepsilon}
\renewcommand{\phi}{\varphi}

\newcommand{\red}[1]{\textcolor{red}{\em #1}}

\newcommand{\xt}{\mathbf{x}}
\newcommand{\yt}{\mathbf{y}}

\newcommand{\kt}{\mathbf{k}}

\newcommand{\ReTr}{\text{Re}\text{Tr}}

\newcommand{\tr}{\text{Tr}}

\newcommand{\V}{V}
\newcommand{\Vd}{V^{\dagger}}

\newcommand{\id}{\mathbb{I}}

\def\<{\langle}
\def\>{\rangle}

\def\n{\nonumber}

\def\l{\left}
\def\r{\right}

\def\f{\frac}

\def\nf{\nicefrac}
\def\tr{{\rm Tr}}

\def\as{{\alpha_s}}

\def\N{\mathbb{N}}

\def\1{\mathbb{1}}
\def\N{\mathcal{N}}
\def\A{\mathcal{A}}
\def\F{\mathcal{F}}

\def\T{\mathcal{T}}

\graphicspath{{graphics/}}

\title{Forward dijet production at the LHC within an impact parameter dependent TMD approach}

\author[1]{Federico Deganutti}
\author[1]{Christophe Royon}
\author[2]{Soeren Schlichting}

\affil[1]{Department of Physics and Astronomy, The University of Kansas, Lawrence KS 66045, USA}
\affil[2]{Fakult\"at f\"ur Physik, Universit\"at Bielefeld, D-33615 Bielefeld, Germany}

\date{}

\begin{document}

\maketitle

\begin{abstract}
We investigate possible signatures of gluon saturation using forward $p+A \to j+j+X$ di-jet production processes at the Large Hadron Collider. In the forward rapidity region, this is a highly asymmetric process where partons with large longitudinal momentum fraction \(x\) in the dilute projectile are used as a probe to resolve the small \(x\) partonic content of the dense target. Such  dilute-dense processes can be described in the factorization framework of Improved Transverse Momentum Distributions (ITMDs). We present a new model for ITMDs where we explicitly introduce the impact parameter (\(b\)) dependence in the ITMDs, to properly account for the nuclear enhancement of gluon saturation effects, and discuss the phenomenological consequences for $p-Pb$, $p-Xe$ and $p-O$ collisions at the LHC. While the case of $p-p$ and $e-p$ collisions is used to fix the model parameters, we find that, on average, the nuclear enhancement of the saturation scale is noticeably weaker than expected from naive scaling with a simple dependence on the atomic number. Since our model explicitly accounts for event-by-event fluctuations of the nuclear geometry, it can also be applied to study forward central correlations in $p-A$ collisions. %in the future.

\end{abstract}

\tableofcontents

\section{Introduction}
\label{sec:introduction}

QCD predictions in the high-energy regime as well as the precise determination of the parton distribution functions (PDFs) have been one of the main goals at the HERA, Tevatron and now LHC colliders.
The observation of a  steep rise of the gluon density in the proton towards small Bjorken-\(x\)  at HERA\cite{H1ZEUS09:InclusiveDISatHera} led to the concept of saturation where
some recombination (non-linear) mechanisms that tames the (linear) growth in gluon multiplicitly should set in at some values of $x$ and $Q^2$ to preserve unitarity. The gluon recombination mechanism is best described in the framework of  Color-Glass Condensate (CGC)\cite{Iancu03:BasicsCGC}, where the linear and non-linear corrections are formalized into the JIMWLK/BK~\cite{Balitsky06:QuarkColorDipoleSmall-xEvolution, kovchegovWeigert06:Small-xTriumvirateRunninCoupling, JalilianKovnerLeonidovWeigert97:WilsonRenormalizationGroupBFKLEquation,
JalialianKovnerWeigert98:WilsonRenormalizationGroupGluonEvolutionFiniteDensity, JalialianKovnerWeigert98:WilsonRenormalizationGroupTowardsHighDensity, IancuLeonidovMcLerran01:NonlinearGluonEvolutionCGC-I, FerreiroIancuLeonidovMcLerran02:NonlinearGluonEvolutionCGC-II} evolution equations.
As a consequence of the interplay of these two competing mechanisms of linear emission and non-linear absorption, a dynamical transverse momentum scale arises (see for instance~\cite{Morreale:2021pnn}).
The understanding of the mechanisms that dynamically give rise to the energy scale of saturation \(Q_s(x)\) is related to dynamics of partons inside hadrons.

Forward di-jet production is one of the preferred processes where the effects of the dynamics of saturation can be investigated at hadron colliders, and has been the subject of numerous theoretical \cite{Marquet:2007vb,Caucal:2021ent,Altinoluk:2023qfr,Boussarie:2021ybe} and phenmenological studies \cite{Albacete:2010pg, LappiMantysaari12:dAuForwardCorrelations,Marquet16:SmallxITMDs, AlbaceteDumitruFujiiNara12:p+PbCGCpredictions, Fujii:2020bkl, Al-Mashad:2022zbq, Gao:2023ulg} (see e.g. \cite{vanHameren:2023oiq} for recent reviews).
% Forward di-jet production is one of the preferred processes where the effects of the dynamics of saturation can be investigated at hadron colliders, and has been the subject of numerous theoretical and phenmenological studies \cite{LappiMantysaari12:dAuForwardCorrelations,Marquet16:SmallxITMDs, AlbaceteDumitruFujiiNara12:p+PbCGCpredictions}.
The detection of jets in the very forward region in the same hemisphere of a typical detector at the LHC imposes a dilute-dense asymmetry where the interacting parton in one of the (dilute) hadrons  carries a large (to intermediate) fraction \(x\) of the momentum of the hadron and is used to probe the  (dense) gluon distribution of the other hadron at much smaller values of \(x\) (see \reffig{geometry} in Sec.~\ref{sec:cross-section} for an illustration of such a configuration).
The process is sensitive to saturation effects when the transverse energy of the small-\(x\) gluon, which is controlled by the transverse momentum imbalance between the two jets \(k_r\), is comparable to the saturation scale.
The interested reader can also find di-jet production analyses dedicated to saturation physics in the context of DIS at e.g.~\cite{Mantysaari:2019hkq, Boussarie:2021ybe, Caucal:2023fsf}.

In this paper, we will compute forward di-jet production cross-section using the unified formalism given in Ref.~\cite{KotkoKutakMarquet15:DiluteDenseImprovedTMDs} that allows to interpolate between the high-energy factorization (HEF) framework\cite{CataniCiafaloniHautmann91:KtFactoriztionTheorem} valid in the regime where the momentum imbalance $k_t$ of the jets is of the order of the transverse momentum $p_j$ of the jets \(k_t\simeq p_j\) and the \emph{Transverse-Momentum-Distribution} (TMD) framework\cite{BomhofMuldersPijlman06:GaugeLinksConstruction, Collaboration23:TMDhandbook}
 valid when the momentum imbalance of the two jets is on the order of the saturation scale \(k_t\lesssim Q_s\).
This Improved-TMD (ITMD) framework has been applied to obtain predictions for forward dijet production at high rapidity in \(p-p\) and \(p-A\) collisions\cite{Marquet16:SmallxITMDs}.
The authors argued that the larger nuclear density in large nuclei results in an earlier onset of the non-linear regime in \(p-Pb\) versus \(p-p\) collisions as the jet geometry approaches the ``back-to-back'' configuration  $(k_t \ll p_j)$.
A significant reduction in jet azimuthal correlation appears in \(p-A\) versus \(p-p\) collisions.
The nucleus distribution was extrapolated from the proton distribution by simply imposing the expected mass number \(A\) scaling, thus treating the nucleus as a uniform object resembling a large proton.

In this paper, we discuss a more accurate approach to describe the nucleus as an incoherent collection of nucleons.
In order to account for these effects on the saturation dynamics, we developed a simple model in which the nuclear TMDs can be computed from the proton TMDs, by promoting the distributions to take into account the impact parameter dependence, given the spatial distribution of nucleons inside the nuclear target.

The starting point of our model is an accurate description of the nuclear matter distribution at small-\(x\).
DIS data at HERA are used as the main source of information for the proton structure because of the high precision of the measurement\cite{H1ZEUS09:InclusiveDISatHera, H1ZEUS18:CharmBeutyProductionHera, H1ZEUS15:InclusiveDISatHera} and of the simplicity of the theoretical interpretation in terms of the dipole picture\cite{NikolaevZakharov91:NuclearShadowingColorTransparencyDIS, NikolaevZakharov91:NuclearShadowingScalingPropertiesDIS, Mueller90:Small-xSaturationModel, Mueller94:SoftGuonsInfiniteMomentumWaveFunction, Mueller94:SoftGuonsInfiniteMomentumWaveFunction, MuellerPatel94:SingleDoubleBFKL-PomeronDipolePicture}. In the dipole picture, DIS cross sections are written in a factorized form that separates the transition of a (virtual) photon to a quark-antiquark ($q \bar{q}$) dipole from the quark-antiquark dipole scattering off the dense nuclear matter in the proton target.
The nuclear matter distribution cannot follow from a pure perturbative description. The small-\(x\) evolution can be computed using perturbative techniques but the nuclear configuration at the starting scale of the evolution must be extracted from experimental data.

In our model we include the small-\(x\) evolution following the procedure of Ref.~\cite{Albacete10:AAMQS}.
The authors (AAMQS) provide a good fit of the most recent HERA data\cite{H1ZEUS09:InclusiveDISatHera} including \(c\) and \(b\) heavy quark contents, where the small-$x$ evolution is driven by the BK equation supplemented with a  subset of the next-to-leading corrections to account for the running of the coupling (rcBK).
Such a choice represents a good compromise between the simplicity of its implementation and its accuracy. A more rigorous treatment would entail a renormalization in terms of the BK-JIMWLK equations which, unfortunately, are notorious for being hard to implement numerically. The performance of the fit gives further confirmation of the validity of this approach.

The AAMQS fit involves only quantities averaged over impact parameter since it is based on the HERA measurements that were not dependent on the centrality of the scattering.
A rigorous treatment of the impact parameter dependence of the dipole amplitude requires the solution of the rcBK equation beyond the translational invariant approximation\cite{AlbaceteKovchegov07:RunningCouplingHighEnergySolution, IkedaMcLerran04:ImpactParameterDependentBK, LappiMantysaari13:SingleProduction}. Such study is complicated by non-perturbative effects that appear at large impact parameters where the onset of the physics of confinement emerges.
Even though a satisfactory treatment of the impact parameter dependent evolution remains elusive, several attempts\cite{KowalskiTeaney03:IPSatModel, RezaeianSiddikovKlundertVenugopalan13:RevisedIpSat, WattKowalski08:ImpactParameterCGC} have been made to introduce the impact parameter dependence into a consistent physical description without having to meddle with the intricacies of the \(b\)-dependent evolution equations.

One prominent example is the Impact Parameter Saturation (IPSat) model~\cite{KowalskiTeaney03:IPSatModel}, which is based on the particularly simple Golec-Biernat and Wusthoff (GBW)\cite{Golec-BiernatWusthoff99:SaturationDiffractiveDIS} dipole saturation model,  but contains on one hand,  a DGLAP extension towards higher \(Q\)\cite{BartelsGolecKowalsky02:DGLAP-SaturationModel} and, on the other hand, an impact parameter dependence in the dipole dynamics. A key advantage of the IPSat model is the fact that it is immediately generalizable from protons to nuclei, such that the model has frequently been employed in studies of $e-A$, $p-A$ and even $A-A$ collisons\cite{KowalskiMotykaWatt06:DiffractiveDipolePictureHERA, KutakSapeta12:LHC-GluonSaturation-pPb, Venugopalan12:IPSatModelVsHera, MantysaariZurita18:HeraDataAnalysisWithWithoutSaturation}. On the other hand, as originally noted by the authors, an obvious shortcoming is the model inability to describe the lower end of the \(x\) evolution since the gluon structure function is only evolved through the (lowest order) DGLAP evolution kernel.

Our model which was inspired by the success of the IPSat model and of the AAMQS fit with rcBK evolution can be seen as an hybrid description combining the two.
In short, we propose to keep the impact parameter dependence of the dipole amplitude \(N\) in the eikonal exponentiated form as done in the IPSat model but we substitute the DGLAP evolution of IPSat with the rcBK evolution extracted from the AAMQS fit.
In this way, the dipole amplitude can then be expected to be more reliable toward small-\(x\) while at the same time,
capturing some of the effects that the \(b\)-averaged AAMQS description lacks.  Since in our study the (transverse) geometry of the nuclei is modeled using the rather sophisticated MCGlauber model~\cite{Miller07:ReviewGlauberModel} that is frequently used in heavy-ion physics,
one can readily anticipate that the model predictions for nuclei deviate significantly from the ``naive'' scaling with the atomic number. Notably, this more detailed treatment will also allow for future extensions of the model to include e.g. centrality dependence or study correlations between forward di-jet production and event activity in the central rapidity region.

The structure of the paper is divided in four sections. Section II starts with a summary of the AAMQS fit model and it is also used to set a common physical framework and notations used in the following sections. Then, what follows is dedicated to a description of our model. It presents the impact parameter dependence  and the values of the free parameters of the model. To that end, we compare the dipole TMDs computed from the dipole AAMQS fits and the \(b\)-dependent TMDs calculated within our model, averaged over impact parameter.
Finally, in section III, we use our model to make predictions for di-jet production in \(p-p\) and \(p-A\) collisions as a probe of saturation effects. Concluding remarks are outlined in Sec.~IV.

\section{Model Description of Nuclear TMDs}
\label{sec:model}

Since our model was designed with the goal of computing the di-jet production cross section in the asymmetric dilute-dense regime in \(p-p\) and \(p-A\) collisions, the task at hand is to infer the various nuclear TMDs that enter the calculation of the ITMD factorized cross-section of \refeq{cross-section} (see also~\cite{KotkoKutakMarquet15:DiluteDenseImprovedTMDs}). The main interest in these processes derives from the fact that they represent an almost ideal probe to study the onset of saturation at hadron colliders. Since the saturation scale is expected to increase with the atomic number \(A\) as \(Q^A_{s}\simeq Q_{s}A^{\nicefrac{1}{6}}\), it has often been argued that the measurement of the di-jet distributions in \(p-A\) and \(p-p\) collisions will be sensitive to the  differences in saturation scales~\cite{Marquet07:AzimuthalCorrelationForwardInclusiveDiJets}. However, in order to make precise predictions, one needs to
properly model the distribution of nuclear matter inside the nucleus that induces the rise of the saturation scale in nuclei.
The challenge is thus to find a consistent way to generalize the available knowledge for protons to arbitrary nuclei.

Our general strategy for tackling this problem can be briefly summarized as follows. Starting point of discussion are the HERA measurements of the proton structure functions \(F_2,F_L\), which can be used~\cite{Albacete10:AAMQS, Venugopalan12:IPSatModelVsHera, MantysaariZurita18:HeraDataAnalysisWithWithoutSaturation} to constrain the (average) dipole amplitude of the proton down to very low values of \(x\). We obtain this information about the average dipole amplitude $\mathcal{N}(x,\vec{r})$ from the AAMQS fit and subsequently generalize the model to account for impact parameter $\vec{b}$ dependence, by introducing a dependence on the local thickness $T_{p}(\vec{b})$. Based on this procedure, we obtain an expression for the dipole amplitude $\mathcal{N}(x,\vec{r},\vec{b})$, which can simply be generalized from protons to nuclei, by replacing the local thickness of the proton $T_{p}(\vec{b})$ with the nuclear thickness $T_{A}(\vec{b})$ obtained as the sum of the thickness of all nucleons inside a nucleus $T_{A}(\vec{b})=\sum_{i \in A}T_{i}(\vec{b})$.
By assuming Gaussian fluctuations of color fields inside the nuclei, we then derive a set of relations, whereby the various TMDs relevant for forward di-jet production can all be expressed in terms of the dipole amplitude $\mathcal{N}(x,\vec{r},\vec{b})$. Since this calculation is technically rather involved, we defer the details to   Appendix~\ref{sec:tmd-calculation}, and simply note that our results generalize earlier results of Ref.~\cite{MarquetPetreskaRoiesnel16:TMDsFromJIMWLK} for some of the TMDs. Based on these relations, we can then perform a sampling of the spatial distributions of nucleons inside nuclei using the MC Glauber model, and calculate the $x$ and $k_t$ dependence of the various TMDs by integrating over the impact parameter $\vec{b}$ and averaging over the spatial distribution of nucleons. Details on the individual steps are provided in the following subsections followed by a brief discussion of our results for the relevant TMDs for proton and nuclear targets.

\subsection{Deep-inelastic scattering (DIS) in the Dipole Picture \& AAMQS fit to HERA data}
\label{sec:dipole-picture}

In the dipole picture used to describe DIS physics at small-\(x\), the interactions between the high-energy photon and the target nuclear matter (the proton) are described in terms of the light-cone wave function \(\Psi^f_{T,L}\), which quantifies the probability for a virtual photon to fluctuate into a quark-antiquark dipole pair of flavor \(f\), and the dipole amplitude \(\N\), which describes the interaction of a $q\bar{q}$ color dipole with the hadronic target. The longitudinal and transverse components of the cross section can be given in the factorized form
\begin{equation}
  \eq{dipole-pic}
  \sigma_{T,L}\lr{x,Q^2}=2\sum_f\int_0^1\!\dd z\intd^2\vec{b}\dd^2\vec{r}\l|\Psi^f_{L,T}\lr{e_f,m_f,z,Q^2,\vec{r}}\r|^2\N(\vec{b},\vec{r},x),
\end{equation}
where \(m_f,e_f\) are the mass and charge of the quark flavors in the dipole,
\(Q^2\) the photon virtuality, \(z\) the longitudinal momentum fraction carried by the quark (see \reffig{dipole}),
\(\N(\vec{b},\vec{r},x)\)  the imaginary part of the dipole-target amplitude that describes the scattering of the dipole off the target, \(\vec{r}\)  the size and position of the dipole, \(\vec{b}\)  the impact parameter that indicates the centrality of the dipole-target collision \footnote{Throughout the paper bold fonts indicate euclidean 2-dimensional transverse vectors.}.
\begin{figure}[bh]
  \begin{center}
    \includegraphics[width=0.3\linewidth]{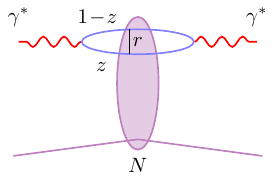}\hspace{10pt}
    \vspace{-10pt}
    \caption{Schematic of DIS in the dipole picture.}
  \fig{dipole}
  \end{center}
\end{figure}

Generally speaking, models of saturation physics aim to extract the behavior of the dipole amplitude \(\N(\vec{b},\vec{r},x)\) from measurements of the inclusive cross-section (and in some cases more exclusive processes~\cite{KowalskiMotykaWatt06:DiffractiveDipolePictureHERA,RezaeianSiddikovKlundertVenugopalan13:RevisedIpSat}) at HERA. Since our calculation of small $x$ TMDs will be based on the extraction of the dipole amplitude of Ref.~\cite{Albacete10:AAMQS} (AAMQS), we will now briefly discuss the simplifications made in the AAMQS analysis. Since the AAMQS analysis was based on inclusive DIS, only impact parameter averaged quantities  are considered.
Introducing (half) the average of the quark distribution in the transverse plane \(\sigma_p\), the cross section in \refeq{dipole-pic} is expressed in terms of the averaged dipole amplitude \(\N(\vec{r},x)\) through the substitution
\begin{equation}
  \label{eq:average}
  2\intd^2\vec{b}\,\N(\vec{b},\vec{r},x) \to\sigma_p \N(\vec{r},x)\;.
\end{equation}
where \(\sigma_p\) was left free in the AAMQS fit analysis and led to the value \(3.54{\rm fm}^2\) for the light quark data set. We keep it fixed to the same value in our analysis.
For simplicity, the heavy quark (charm and bottom) data set has been excluded from our analysis.

Within the analysis, the averaged dipole amplitude \(\N(x,\vec{r})\) was evolved starting from a non-perturbative initial condition modeled following the semi-classical  Mueller-Venugopalan model \(MV^{\gamma}\)~\cite{MvLerranVenugopalan94:ColorFieldGreenFunctions, MvLerranVenugopalan93:Small-k_t-GluonDistribution, MvLerranVenugopalan93:QuarkGluonDistributions}
\begin{equation}
  \label{eq:MVgamma}
  \N(x=0.01,\abs{\vec{r}})=
1-\exp\l[-\f{\lr{r^2Q^2_{s0}}^\gamma}{4}\log\lr{\f{1}{r\Lambda}+e}\r]\;,
\end{equation}
where \(Q_{s0},\Lambda,\gamma\) are additional fitted parameters and \(e\) is the Euler constant. \refeq{MVgamma} gives the ``exponentiated'' form of the dipole amplitude originating from the multiple rescatterings
of the dipole passing through a dense target shock-wave\cite{KowalskiTeaney03:IPSatModel}.
Starting from the initial scale $x=0.01$, the evolution towards smaller \(x\) was performed using the Balintsky-Kovchegov (BK) equation\cite{Balitsky06:QuarkColorDipoleSmall-xEvolution, kovchegovWeigert06:Small-xTriumvirateRunninCoupling} with running coupling corrections (rcBK)\cite{AlbaceteKovchegov07:RunningCouplingHighEnergySolution, AlbaceteArmestoMilhanoSalgado05:NuclearSizeRapidityDependenceSaturationScale}. This evolution represents the dynamical input based on the QCD perturbative predictions.
The explicit form of the rcBK integral-differential equation can be found in Eqs.~(2.7, 2.8) of Ref.~\cite{Albacete10:AAMQS}.

We use the FORTRAN routine provided  in Ref.~\cite{Albacete10:AAMQS} to compute the dipole amplitude \(\N(x,\abs{\vec{r}})\) and generate a data set mapping for any chosen value of \(x\lr{\in(10^{-12},0.008}\) of the dipole sizes \(r=\abs{\vec{r}}\) with the corresponding dipole distributions \(\N\lr{r,x}\). The available dipole sizes span several order of magnitudes \(r_i\in\lr{10^{-8},10}\).

Since the calculation of the TMDs involves taking derivatives over \(\vec{r}\),
a function \(\N^*(r,x_n)\), where the subscript \(n(=1\dots 8)\) enumerates the \(x\) sampled values was fitted to each data set  corresponding to a fixed value of \(x=x_n\). The choice of the function \(\N^*\) used for the fit was inspired by the \(MV^\gamma\) form of the dipole amplitude\footnote{Note that  the goal is to reproduce the AAMSQ fit in the most general way. The possibility of over fitting the data with so many parameters is of no concern since we do not extrapolate the fit beyond the \(x\) range of the measurement and we just use the fit to reproduce data. We do not claim a particular physical interpretation for that parametrization.}
\begin{subequations}\label{eq:dipole-mod}
\begin{align}
  \N^*\lr{r,x}&=1-\exp\l[-G\lr{r,x}\r]\;, \label{eq:dipole-mod-a}\\
  G\lr{\tilde{r},x=x_n}&=\f{\lr{\tilde{r}^2q^2\lr{x_n}}}{4}^{\gamma\lr{x_n}}\!\!\!\!\!\!\!\!\log\lr{\f{1}{\tilde{r}\lambda\lr{x_n}}+e e_c\lr{x_n}}\;. \label{eq:dipole-mod-b}
\end{align}
\end{subequations}
We note that in the above expressions, the dipole size \(r\) is rescaled as \(\tilde{r}=r/r_s\), where \(r_s\) is the inverse of the saturation scale, i.e.
\begin{align}
   Q_s(x_n)=1/r_s(x_n)\;,
\end{align}
which is computed for each data set, by following the standard procedure in solving the implicit equation
\[G\lr{r_s,x_n}=\f{1}{4}.\] By interpolating the results based on this parametrization, the exponent of the dipole amplitude \(G(\tilde{r})=G\{{q}_n,\lambda_n,\gamma_n,{e_c}_n\}(\tilde{r})\) is then given as a smooth function of \(\tilde{r}\) that depends on a family of dimensionless parameters~\footnote{The parameters \(Q_{s0}\) and \(\Lambda\) in the usual \(MV^\gamma\) expression (see \refeq{MVgamma}) were renamed to \(q\) and \(\lambda\) respectively to emphasize the change in their meaning when using the dimensionless variable \(\tilde{r}\) .}.
We note  that the quality of the fit in Eqs.~\ref{eq:dipole-mod} remains within \(10\%\) of data throughout the full range of dipole sizes.

\subsection{Impact Parameter Dependence}

Evidently, in the proton case, the fit to HERA data is of good quality which means that the  physics of inclusive DIS can be well described without taking into account any impact parameter dependence.
However, we expect the impact parameter dependence to play an important role in the
context of \(p-A\) collisions. In nuclei, the nuclear densities can vary greatly according to the nucleon positions resulting in an irregular shape with lumps and valleys whose effects are hard to capture with a simple scaling correction factor. Hence, we will now discuss how to  introduce the \(b\)-dependence into the description starting from the \(b\)-averaged rcBK fit.

Inspired by the simple but successful IPSat Model\cite{KowalskiTeaney03:IPSatModel}, we include a proportionality to the nuclear matter density in the exponential, as it is done for the Glauber-Mueller formula\cite{Mueller90:Small-xSaturationModel} for the dipole cross section. Starting from \refeq{dipole-mod}, we thus impose the following substitution
\be
\label{eq:IPmodel}
  \N^*(\vec{r},x)\to\N^*(\vec{b},\vec{r},x)=
  \l\{
  \begin{aligned}
    &\eta \f{S^p_\perp}{\sigma}\lr{1-\exp\l[-\sigma T(\vec{b})G(r,x)\r]}\ &T\geq T_{\rm cut}\;, \\
    &0 &T<T_{\rm cut}\;,
  \end{aligned}
  \r.
\ee
where the nuclear thickness \(T(\vec{b})\) is the two-dimensional projection of the nuclear density \(\rho(z,\vec{b})\) over the transverse impact parameter space (c.f. Appendix~\ref{sec:nuclear-matter}), \(S^p_\perp=\sigma_p/2\) is the proton transverse area fixed to the AAMQS value. \(\sigma\), \(\eta\) and \(T_{\rm cut}\) are model parameters to be fixed according to the procedure detailed in Sec.~\ref{sec:paramter-fix}. Note that having set an overall \(S^p_\perp\) factor in \refeq{IPmodel} the parameter \(\eta\) is dimensionless while \([\sigma]={\rm fm}^2\).
Specifically, for the proton we follow the IPSat model and employ a Gaussian profile
\[T_p(\abs{\vec{b}})=\f{e^{-\vec{b}^2/\lr{2B_G}}}{2\pi B_G}\;,\]
with a nucleon (square) width \(B_G=4 {\rm GeV}^{-2}\)~\cite{RezaeianSiddikovKlundertVenugopalan13:RevisedIpSat}. When considering nuclei, we simply replace the \(T_p\to T_A\) without modifying the parameters of our model, where the nuclear thickness $T_A$ is obtained as the sum of the thickness functions $T_{p/n}$ of all nucleons\footnote{We make no distinction between proton (\(p\)) and neutron (\(n\)) transverse distributions \(T_p=T_n.\)}. Due to the fluctuating positions of the nucleons in the nucleus, which we sample according to a MC Glauber procedure, as explained in Appendix~\ref{sec:nuclear-matter}, the nuclear thickness fluctuates on an event-by-event basis. The final results for the TMDs are obtained taking the statistical averages of the TMDs computed from typically $\approx 10$ nucleon configurations.

We note that a lower bound \(T_{\rm cut}>0\) on the nuclear density is necessary in \refeq{IPmodel} to recover a finite result once the \(\intd^2\vec{b}\) integral is computed.
Physically, a finite cut-off can be justified since a description in terms of saturation physics is valid only when the dipole interacts with a substantial amount of nuclear matter, and thus has a high probability of subsequent multiple rescatterings. Due to this cut-off, it is necessary to introduce a further dimensionless parameter $\eta$, so  that the $\vec{b}$-dependent model correctly reproduces the average proton TMDs upon integrating over the impact parameter. Since the behavior of $G(r,x)$ is readily fixed by the AAMQS fit, the parameters \(\sigma,\eta\) and \(T_{\rm cut}\) are then the only free parameters of our model, which can be fixed by requiring that the proton TMDs calculated in our model match the proton TMDs computed from the average dipole amplitude in Eqs.~\ref{eq:dipole-mod}.

\subsection{Calculation of small $x$ TMDs}
\label{sec:tmds}
Based on the knowledge of the parametrization of the dipole amplitude in \refeq{IPmodel}, we can proceed with the calculation of the small $x$ TMDs that enter the di-jet production cross section. Since, in the small $x$ limit, the TMDs can be formulated in terms of correlation functions of light-like Wilson lines\cite{CherednikovStefanis08:WilsonLinesTMDs, Marquet:2016cgx, vanHameren:2016ftb}, it is possible to express the various TMDs in terms of the exponent \(G(x,\vec{r},\vec{b}) \) of the fundamental dipole amplitude, if one invokes the additional assumption of Gaussian statistics for the distribution of color charges inside the nucleus, that is common to almost all saturation models~(see e.g. \cite{Lappi:2015vta} for a discussion). Denoting the two end points of the dipole as \(\vec{x},\vec{y}\), such that \[G_{\vec{x},\vec{y}}=G\Big(\vec{r}=\vec{x}-\vec{y},\vec{b}=\frac{\vec{x}+\vec{y}}{2}\Big)
\]
one finds for example, that
the first TMD $\mathcal{F}_{qg}^{(1)}$ in the quark-gluon channel -- which corresponds to the usual fundamental dipole gluon distribution -- can be expressed,  in (transverse) coordinate space, in terms of $G_{\vec{x},\vec{y}}$ and its partial derivatives \(G^{(i,j)}_{\vec{x},\vec{y}}(x)\)~\footnote{Note that we introduce a negative sign in the definition
\[G^{(i,j)}_{\vec{x},\vec{y}}(x)\equiv -\f{\partial^i}{\partial\vec{x}^i}\f{\partial^j}{\partial\vec{y}^j}G^{(i,j)}_{\vec{x},\vec{y}}(x).\]} as
\be
\mathcal{F}_{qg}^{(1)}(\mathbf{x,y})=
e^{G^{(0,0)}_{\mathbf{x,y}}}\left(G_{\mathbf{x,y}}^{(i,i)}-G_{\mathbf{x,y}}^{(i,0)}G_{\mathbf{x,y}}^{(0,i)}\right)\;.\eq{tmd-def}
\ee
up to a factor of \(c=\f{N_c}{\as(2\pi)^2\pi}\), which we omitted from the definition of the TMD in (transverse) coordinate space and also exclude from all plots of TMDs in the following.
Similar expressions can also be derived for all the other TMDs, following the procedure outlined in Appendix~\ref{sec:tmd-calculation}. Since the calculation is rather cumbersome, and the resulting formulae are not particularly enlightning, we refrain from reproducing all the expressions in the main part of this paper, and instead refer the interested reader directly to Appendix~\ref{sec:tmd-calculation}, where we provide the explicit expressions of all TMDs relevant to our analysis.

Based on these expressions, we follow the same logic as in \refeq{IPmodel} and employ
\begin{gather}
G_{\vec{x},\vec{y}}(x)\simeq \sigma T(\vec{b})G({\abs{\vec{r}}},x)\;, \label{eq:Gdipole}
\\
\intertext{to calculate the various TMDs $\mathcal{F}$ in coordinate space, by means of the substitution}
 \F(\vec{r},\vec{b};x)=\F\lr{\{G(\vec{r},\vec{b};x)\}}\simeq \eta \f{S_\perp}{\sigma}\F\lr{\{\sigma T(\vec{b})G(\vec{r};x)\}}   \label{eq:TMD-IPmodel}
\end{gather}
in the respective expression. Subsequently, the TMDs in momentum space are obtained by a simple Fourier transform, i.e.
\begin{subequations}
  \begin{align}
  \eq{bessel}
F(x,\vec{k})&=c \intd^2\vec{b}\int\!\dd^2\vec{r} e^{-i\vec{k}\cdot\vec{r}}\mathcal{F}(x,\vec{r},\vec{b})\;,
  \end{align}
  where the transverse momentum \(\vec{k}\) is the Fourier conjugated variable to the dipole size \(\vec{r}\). Since, for unpolarized nuclei, the distributions are  azimuthaly symmetric, this amounts in practice to calculate the Bessel integral
  \begin{align}
  F(x,k_t)&=c \intd^2\vec{b}\intd r~2\pi r~J_0(k_t r)\mathcal{F}(x,r,\vec{b})\;,
  \end{align}
  \end{subequations}
which we obtain by means of numerical integration.

\subsection{Fixing the Model Parameters}
\label{sec:paramter-fix}

\begin{figure}[htbp]
  \begin{center}
    \includegraphics[width=0.8\textwidth]{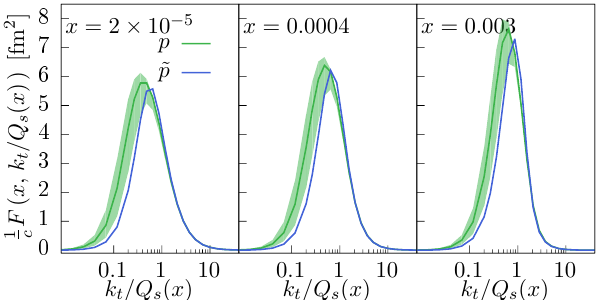}
    \caption{Dipole TMDs (up to a constant $c=\frac{N_c}{\alpha_s\pi(2\pi)^3}$) for the proton with (\(p\)) and without (\(\tilde{p}\)) impact parameter (\(\mathbf{b}\)) dependence for three different values of \(x\). Colored band represents the uncertainty due to the cut-off variations in the impact parameter dependent model. The curves where rescaled along the \(\hat{x}-axis\) for convenience of representation. The rescaling \(Q_s(x)\) values are \(\{0.99, 0.65, 0.49\}\) GeV in order of increasing $x$ values. Model predictions reproduce the \(b\)-independent dipole TMD at large \(k_t/Q_s\gtrsim 2-3\) with great accuracy. }%
    \label{fig:TMDmatching}
  \end{center}
\end{figure}

Now that we have established the procedure to calculate the various TMDs, we will discuss how the parameters of our model are fixed by requiring that the \(b\)-dependent proton dipole TMD \(\F^{(1)}_{qg}\) reproduces the rcBK fit once it is averaged over impact parameter space. In the following, the former and the latter will be indicated as $p$ and $\tilde{p}$ respectively, such that the matching condition takes the form
\begin{equation}
  \overline{F}^{(1)}_{qg}(x){\big|_{\tilde{p}}} \stackrel{?}{=} \langle F^{(1)}_{qg}(x,\vec{b}){\big|_{p}}\rangle\;, \label{eq:matching}
  \end{equation}
  where
\begin{subequations}
\begin{align}
    {\overline{F}^{(1)}_{qg}(x)}{\big|_{\tilde{p}}}
    =&S^p_\perp\,F^{(1)}_{qg}\lr{\{G_{\vec{r}}(x)\}}\;, \\
    \langle {F^{(1)}_{qg}(x,\vec{b})}_{\big|_p}\rangle
    =&\eta\frac{S^p_\perp}{\sigma}\int_{T(\vec{b})>T_{\rm cut}} \!\!\!\!\!\!\!\!\!\!\!\! \dd^2\vec{b} F^{(1)}_{qg}\l(\{\sigma T_p(\vec{b})G_r(x)\}\r)\;. \label{eq:TMDmodel}
\end{align}
\end{subequations}
Based on this requirement, the parameters \(\sigma\) and \(\eta\) can be chosen so that the perturbative large \(k_t\) ``tails'' of the proton dipole TMDs match as shown in \reffig{TMDmatching}. By looking at the explicit expressions for the TMDs
\begin{subequations}
\label{eq:DipoleAVG}
\begin{align}
\frac{{\overline{F}^{(1)}_{qg}}_{|_{\tilde{p}}}}{S^p_\perp}&=\int\!\f{\dd r r}{(2\pi)^2} J_0(k r)
\left[G^{(1,1)}(r,x)-G^{(1,0)}(r,x)G^{(0,1)}(r,x)\right]e^{G^{(0,0)}(r,x)}\,,\\
\begin{split}
\frac{\langle F^{(1)}_{qg}(b)\rangle_{|_p}}{S^p_\perp}&=\int\!\f{\dd{r}r}{(2\pi)^2} J_0(k r)\times\\
&\quad\frac{\eta}{\sigma}\!\!\int_{T(b)\geq T_{cut}}\!\!\!\!\!\!\!\!\!\!\!\! \dd^2\vec{b}\!
\left[\sigma T_p(\vec{b})G^{(1,1)}(r,x)-\sigma^2 T^2_p(\vec{b})G^{(1,0)}(r,x)G^{(0,1)}(r,x)\right]e^{\sigma T_p(\vec{b})G^{(0,0)}(r,x)}\,,
% \\
% &=\int dr\frac{2\pi r}{(2\pi)^3} J_0(k r)\times\\
% &\!\!\int_{T(b)\geq T_{cut}}\!\!\!\!\!\!\!\!\!\!\!\! d^2b\!
% \left( \frac{T_p(b)}{\mathcal{A}_p}G^{(1,1)}(r,x)-\frac{T^2_p(b)}{\mathcal{A}_p\mathcal{T}_p}G^{(1,0)}(r,x)G^{(0,1)}(r,x)\right)e^{ \frac{T_A(b)}{\mathcal{T}_p}G^{(0,0)}(r,x)}
\end{split}
\end{align}
\end{subequations}
one realizes that in the large $k$ limit, the exponentials can be approximated as unity, giving rise to the following matching conditions
 \begin{subequations}
\begin{align}
  \label{eq:parameters}
% \sigma_0=
\eta=&\frac{1}{\mathcal{A}_p}\;:
&\mathcal{A}_p=\int_{T(\vec{b})\geq T_{\rm cut}}\!\!\!\!\!\!\!\!\!\!\!\! \dd^2\vec{b}\ T_p(\vec{b})\;,\qquad\\
% \sigma_1=
\sigma=&\f{1}{\T_p}\;:
 &\A_p\mathcal{T}_p=\int_{T_p(\vec{b})\geq T_{\rm cut}}\!\!\!\!\!\!\!\!\!\!\!\! \dd^2\vec{b}\, T^2_p(\vec{b})\;.\qquad
\end{align}
\end{subequations}
%\end{gather}
%
We note that \(\T_p\) is the average thickness value, while the parameter \(\A_p\lesssim 1\) because of the finite cut-off $T_{\rm cut}$. Based on our discussion, a reasonable value for
 \(T_{\rm cut}\) is provided by \(T_{\rm cut}=T_p(b_{\rm Max})\) with \(b_{\rm Max}\simeq 2\sqrt{B_G}\). Since this choice is to some extent ad-hoc, the model uncertainty was estimated by varying this cut-off value by \(\pm 50\%\).  We note that --- as can be easily understood from their definitions (see \refeq{parameters}) --- the model parameters \(\sigma\) and \(\eta\) must be recomputed when \(T_{\rm cut}\) varies. Their explicit values are collected in Table~\ref{tab:params1}.

We include this model uncertainty in all figures, where the colored bands indicate the uncertainties due to the cut-off value. We verify this procedure in \reffig{TMDmatching}, where we show the dipole TMDs for both the $b$-dependent and $b$-averaged approaches for three different values of $x$. We find that, as can be expected from the matching procedure, the $p$ and $\tilde{p}$ TMDs show small differences at low values of $k_t\sim Q_{s}(x)$, but are in excellent agreement in the perturbative large $k_t\gg Q_{s}(x)$ region.

\begin{figure}[htbp]
\begin{center}
\includegraphics[width=0.95\textwidth]{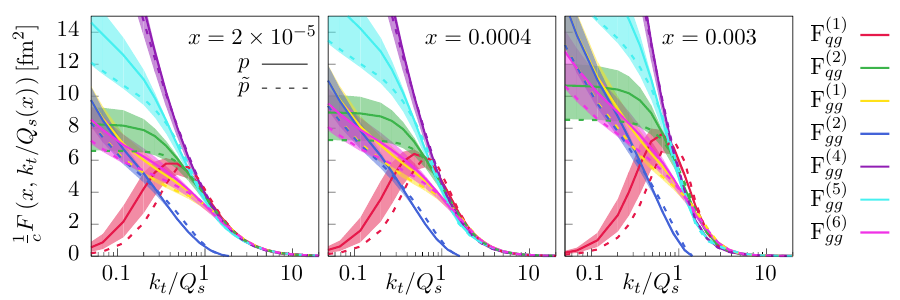}
\caption{Different TMDs (up to a constant $c=\frac{N_c}{\alpha_s\pi(2\pi)^3}$) entering the di-jet production calculation for the proton (\(p\)) for three different values of \(x\). Dashed lines show the \(\mathbf{b}\)-averaged proton cases (\(\tilde{p}\)), while solid lines correspond to the impact parameter dependent model. The colored bands indicate the model uncertainty range.}
\label{fig:all-tmds}
\end{center}
\end{figure}

Similar behavior can also be observed for the other TMDs that enter the calculation of the di-jet production cross-section, which are compactly summarized in \reffig{all-tmds}, where we present the results for the seven relevant proton TMDs.\footnote{Note that the other TMD involved in the calculations\cite{KotkoKutakMarquet15:DiluteDenseImprovedTMDs} is
\[
F^{(4)}_{gg}=F^{(3)}_{gg}\;.\]} Notably --- with the exception of $F^{(2)}_{gg}$ (c.f.\ discussion in~\cite{KotkoKutakMarquet15:DiluteDenseImprovedTMDs}) --- all the TMDs exhibit the same perturbative large $k_t/Q_{s}(x)$ behavior, which is in excellent agreement between the average $\tilde{p}$ and impact parameter dependent $p$ model. Conversely, the low momentum behavior is different between the different TMDs, and it turns out to be important to include the full finite $N_c$ expressions in App.~\ref{sec:tmd-calculation} for the various TMDs, to properly capture the low momentum behavior. While at low momentum $k_t/Q_{s}(x) \lesssim 1$ there are visible differences between the results for the average proton and the impact parameter dependent model, the overall behavior of the two model TMDs is still quite similar. % between two models of the proton.

\begin{table}[htbp]
  \begin{center}
    \begin{tabular}{ |c | c c| }
    % \begin{tabular}{ |c||c|c| }
      % \begin{tabular}{ p{2cm}||p{2cm}|p{2cm} }
      \multicolumn{3}{c}{Model Parameters} \\
      \hline
    \(T_{\rm cut}\,\l[{\rm fm}^{-2}\r]\) & \(\eta\) & \(\sigma\,\l[{\rm fm}^2\r]\)\\
      \hline
      0.1 & 1.11 & 1.78\\
      0.14 & 1.16 & 1.72\\
      0.2 & 1.24 & 1.63\\
      \hline
    \end{tabular}
    \caption{\(\sigma\) and \(\eta\) values for different values of \(T_{\rm cut}\).}
    \label{tab:params1}
  \end{center}
\end{table}

\subsection{Model predictions for the proton \(F_2\) structure function measured at HERA}

Before we proceed to the calculation of nuclear TMDs, we perform a closure test on the $b$-dependent model by comparing its predictions for the DIS proton structure function \(F_2\) to the HERA data used in the AAMQS fit. The structure function \(F_2\) is given in terms of the dipole amplitude \(\N\) as
\begin{equation}
F_2(Q^2,x,\N)=\frac{Q^2}{4\pi^2\alpha_{\rm em}} 2\sum_f\sum_{\chi=L,T}\intd z\intd^2\vec{r}\abs{\Psi^f_\chi(Q^2,r,z)}^2\N\;,
\end{equation}
where
\(\Psi_{L,T}(Q^2,r,z)\) are the longitudinal (L) and transverse (T) photon wave functions.
Their explicit expressions can be found at Ref.~\cite{Golec-BiernatWusthoff98:SaturationEffectsDISDiffraction}.
Thus, we compare
\begin{equation}
  F_2(Q^2,x,\N(r,x))\stackrel{?}{=}\langle F_2(Q^2,x,\N^*(r,\vec{b},x))\rangle
  \equiv\intd^2\vec{b}F_2(Q^2,x,\N^*(r,\vec{b},x))
 \end{equation}
where \(\N\) and \(\N^*\) are coming from the AAMQS fits (see \refeq{dipole-mod}) and the model dipole amplitudes (see \refeq{IPmodel}) respectively.

\begin{figure}[htbp]
  \begin{center}
  \includegraphics[width=0.85\textwidth]{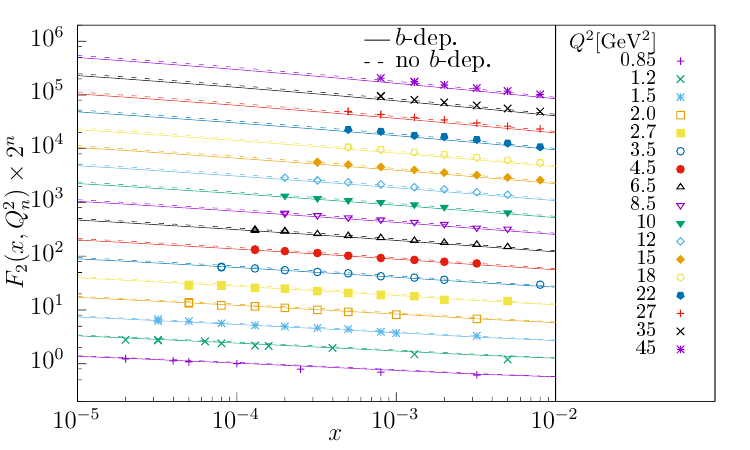}
  \caption{\(F_2(x,Q^2)\) (dashed line) and \(\langle F_2(x,Q^2,\mathbf{b})\rangle\) (full line) compared to HERA data.
The plots corresponding to different photon virtualities \(Q^2=Q^2_n\) are shifted along the \(y\)-axis by a factor of \(2^n\), where $n=0$ to 16 enumerates the \(Q^2_n\) values starting from \(Q^2_{n=0}=0.85\) GeV\textsuperscript{2}, to avoid the overlapping between the different curves. We note a good agreement with data using both approaches, which validates our model for the $b$ dependence.}%
\label{fig:F2}
\end{center}
\end{figure}

The fit quality depicted in \reffig{F2}, which is comparable using both parametrizations ($\N$ and $\N^*$) supports the validity of our model for the $b$-dependence.

\subsection{Model predictions for nuclear TMDs}

\begin{figure}[htbp]
\begin{center}
\raisebox{-0.5\height}{\includegraphics[width=0.59\textwidth]{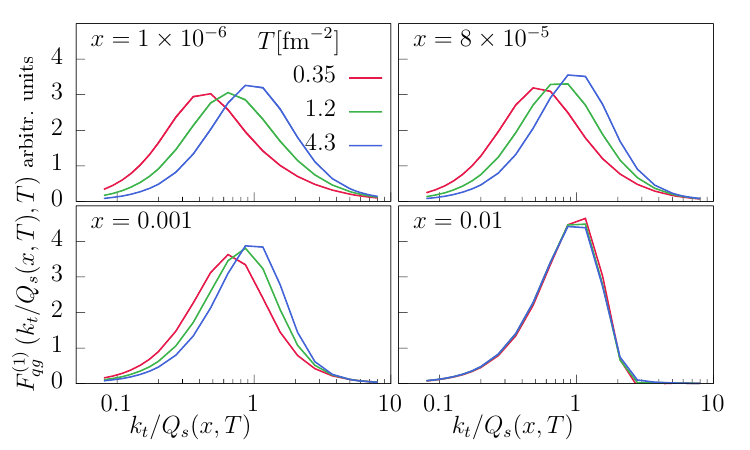}}
\hspace{-10pt}
\raisebox{-0.5\height}{\includegraphics[width=0.4\textwidth]{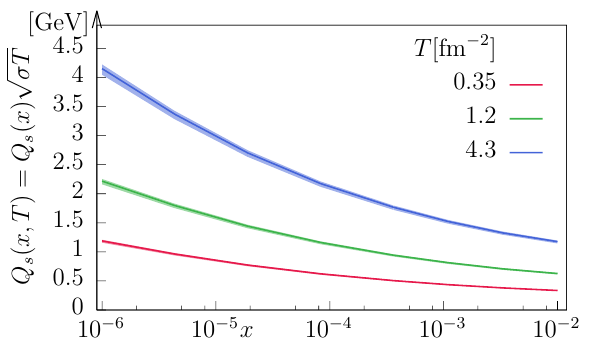}}
\caption{Left: (Unintegrated) \(F^{(1)}_{qg}\) TMDs plotted as a function of the (rescaled) transverse momentum \(\tilde{k}_t=k_t/Q_s(x,T)\) for small, intermediate and large values of the nuclear thicknesses for four different values of \(x\). The line thickness indicates the uncertainty due to the choice of the cut-off value. Right: Saturation scale for three different values of $T$ as a function of $x$.}
\label{fig:tab-tmds}
\end{center}
\end{figure}

Now that we have established the procedure for calculating small-x TMDs, we proceed to the calculation of the nuclear TMDs. We note that in order to perform an efficient computation of the TMDs, which as shown in \refeq{bessel} involve numerical integrals of Bessel functions that demand a large number of integrand evaluations, it is convenient to tabulate the TMDs as a function of the  nuclear thickness value \(T\). One example of these unintegrated (w.r.t. \(\vec{b}\)) TMDs is shown in the left panel of \reffig{tab-tmds}, where we present the unintegrated dipole TMD \(F(x,T=T(\vec{b}))\) as a function of the rescaled momentum \(k_t/Q_s(x,T)\). Here \(Q_s(x,T)=Q_s(x)\sqrt{\sigma T}\) denotes the saturation scale for different nuclear thicknesses, which is depicted in the right panel of \reffig{tab-tmds} as a function of $x$ for three values of $T$. While at larger values of $x$, the dependence on the nuclear thickness $T$ can almost entirely be accounted for by use of the scaling variable $k_t/Q_s(x,T)$, a more non-trivial $T$ dependence emerges at smaller values $x$ following the rcBK evolution.

\begin{figure}[htbp]
\begin{center}
\raisebox{-0.5\height}{\includegraphics[width=0.6\textwidth]{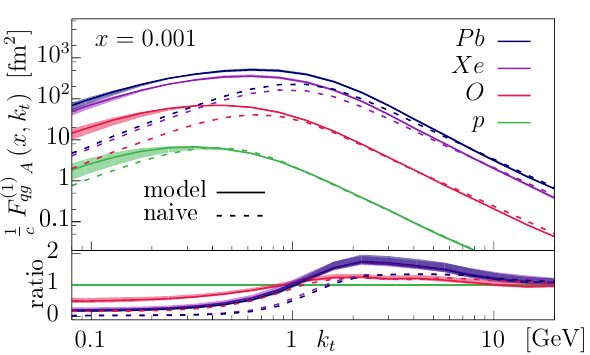}}
   \hspace{-15pt} \raisebox{-0.5\height}{\includegraphics[width=0.39\textwidth]{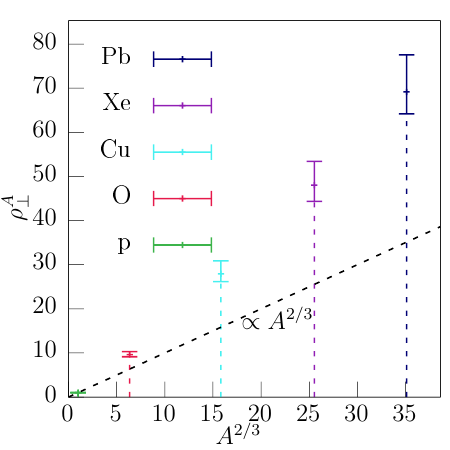}}
    \raisebox{-0.5\height}{\includegraphics[width=0.85\textwidth]{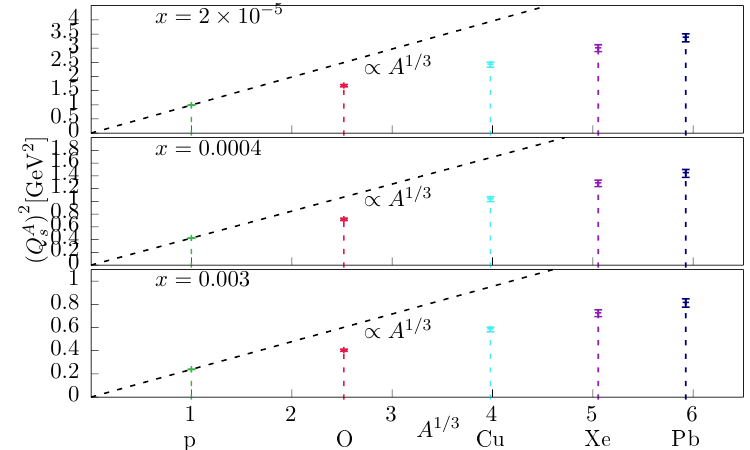}}
\caption{Top Left: \({F^{(1)}_{qg}}_A(k_t,x=0.001)\) TMD (up to a constant $c=\frac{N_c}{\alpha_s\pi(2\pi)^3}$) for proton (\(p\), green), Oxygen (\(O\), red) and Lead (\(Pb\), blue) in ``naive'' (dashed line) and model (full line) predictions as a function of \(k_t\) measured in GeV. Absolute values are shown in the top plot while the nuclei \emph{versus} proton TMD ratios (normalized to $A$) are shown at the bottom. The colored bands indicate the model uncertainties.
Top Right: TMD magnitude (\(\hat{y}-axis\)) rescaling factor \(\rho^A_\perp\) for proton and all nuclei as a function of \(A^{2/3}\). The model predicts a nuclear mass number dependence stronger then the naive \(\sim A^{2/3}\) expectation (dashed line). The error bars indicate the model uncertainty. We note that $\rho^{A=p}_\perp=1$.
Bottom: Saturation scales (squared) for proton, Oxygen, Copper, Xenon and Lead as a function of \(A^{1/3}\) and the respective model uncertainties indicated as colored error bars. The dashed line shows the nucleus saturation scales in the case of uniform scaling \(Q^2_s(A)\sim A^{1/3}Q^p_s\). Our model predicts a weaker dependence on the atomic mass number of the nucleus saturation scale. }
\label{fig:NuclearTMDSummary}
\end{center}
\end{figure}

Knowing the un-integrated TMDs \(F(x,T=T(\vec{b}))\), the integral over impact parameters in \refeq{bessel} can then simply be carried out by mapping the value of \(\vec{b}\) to the corresponding unintegrated TMD \(F(x,T=T(\vec{b}))\). The TMDs shown in all other figures that will be discussed in the following have already been integrated over \(\vec{b}\).  Once again, we first present results for the dipole TMD, which is shown in the top left panel of Fig.~\ref{fig:NuclearTMDSummary} for proton (p), Oxygen ($O_{16}$), Xenon (\(Xe_{129}\)) and Lead $(Pb_{208}$) nuclei. Solid bands in Fig.~\ref{fig:NuclearTMDSummary} show the results obtained from the impact parameter dependent model, while dashes lines correspond to the naive scalings of $F \propto A^{2/3}$ and $k_t^2 \propto A^{1/3}$ with the atomic mass number $A$. Generally, we find that the uncertainties of the impact parameter dependent model appear to be well under control, and the results show the expected behavior that the dipole TMD of a nucleus is larger due to the increase of the transverse area and peaks at larger values of the transverse momentum $k_t$ due to the increase of the saturation scale. However, the naive scaling with atomic number does not properly account for this dependence, as can be seen from the sizeable deviations from the impact parameter dependent predictions for all nuclei. We further elaborate on this behavior in the top right and bottom panel Fig.~\ref{fig:NuclearTMDSummary}, where we present the $A$ dependence of the saturation scale
\begin{align}
  Q^A_s(x)\equiv&Q_s(x)\sqrt{\sigma \T_A}\;,\\
  \intertext{at three different values of $x$ and of the nuclear TMD rescaling ratio}
  \rho^{A}_\perp=&S^{A}_\perp/S^{p}_\perp\;,\eq{rho}\\
  % =\f{\T_p}{\T_A}\f{\A_A}{\A_p}
  \shortintertext{where}
  S^{X=A,p}_\perp=&\f{\sigma_p}{2} \f{\A_X}{\T_X}\frac{\eta}{\sigma}\;,
% S^{p/A}_\perp=\intd^2\vec{b}\,\theta\lr{T_{p/A}(\vec{b})-T_{\rm cut}}\;.
\end{align}
estimates the effective transverse areas of protons (\(S^p_\perp\)) or nuclei (\(S^A_\perp\)) in the model.
The \(\A_A\) and \(\T_p\) values for each nucleus are computed analogously to  \(\A_p\) and \(\T_p\) of \refeq{DipoleAVG} and are reported in Table~\ref{tab:rescaling}.

When considering the results in Fig.~\ref{fig:NuclearTMDSummary}, we find that as a function of the atomic number the saturation momentum $\big(Q_{s}^{A}(x)\big)^2$ increase more slowly than the naive $A^{1/3}$ scaling, whereas the effective transverse area increases more rapidly than the naive $A^{2/3}$ power law\footnote{Note that such opposite tendencies for the saturation scale and effective transverse area with respect to the naive expectations are a necessary requirement in order for the (approximate) sum rule  \[\l.\intd^2\vec{k} F_A(\vec{k},x)\middle/\!\!\intd^2\vec{k}F_p(\vec{k},x)\simeq A\r.\;\]
to be fulfilled.}.
Intuitively, this can be understood from the fact that a large part of the cross section is dominated by scattering off the relatively large edges of the nucleus, where the nuclear matter densities do not significantly exceed those inside a proton, such that the nuclear enhancement of saturation effects is rather mild. We thus conclude that a proper treatment of the impact parameter dependence is important in order to make accurate saturation physics predictions for nuclei.

\begin{figure}[htbp]
\begin{center}
 \raisebox{-0.5\height}{\includegraphics[width=0.9\textwidth]{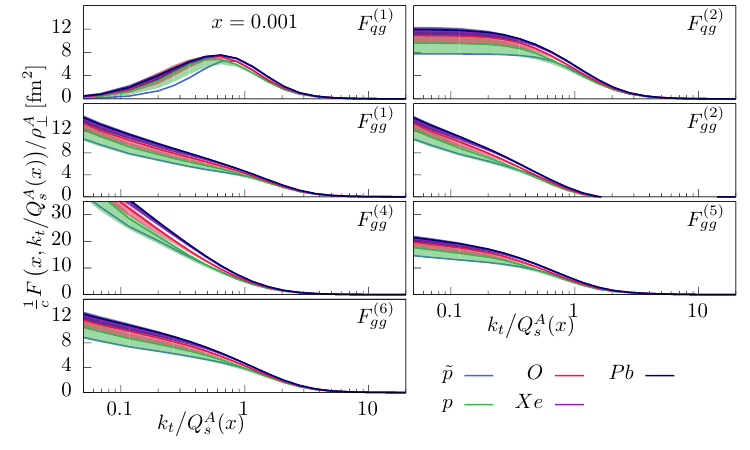}}
\raisebox{-0.5\height}{\includegraphics[width=0.49\textwidth]{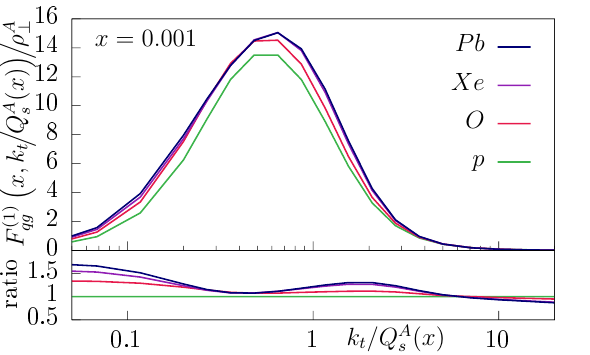}}
\hspace{-10pt}  \raisebox{-0.5\height}{\includegraphics[width=0.49\textwidth]{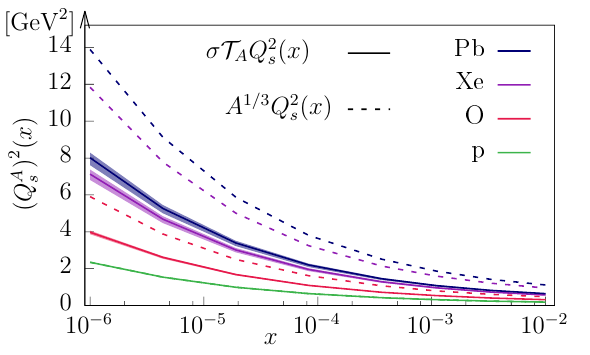}}
\caption{Top: All 7 independent (rescaled) TMDs (\(F^{(3)}_{gg}=F^{(4)}_{gg}\)) (up to a constant $c=\frac{N_c}{\alpha_s\pi(2\pi)^3}$) at \(x=0.001\) for \(b\)-dependent (\(p\)) and \(b\)-independent (\(\tilde{p}\)) proton, Oxygen (\(O\)), Xenon (\(Xe\)) and Lead (\(Pb\)). The colored bands represent the uncertainty due to the cut-off variations.
Bottom Left: Model predictions for proton, Oxygen, Xenon and Lead dipole TMDs \(F^{(1)}_{qg}\) (up to a constant $c=\frac{N_c}{\alpha_s\pi(2\pi)^3}$). The bottom plot shows the result for different ions relative to the proton. At intermediate to large  transverse momenta the (rescaled) TMDs match exactly whereas at smaller values the ratio can (slightly) exceeds values of 1.5 at about \({k}_t/Q^A_s\lesssim 0.1\).
Bottom Right:  Nucleus squared saturation scale as a function of $x$.
The saturation scale (squared) from our model (shown in full line) grows slower than the ``naive''\(\big(Q^A_s(x)\big)^2\propto Q^2_s(x)A^{1/3}\) expectation (shown in dashed line).
By construction, the model prediction for the  proton saturation scale matches exactly the ``average'' proton case \(Q^p_s\equiv\sqrt{\sigma\T_p}Q_s=Q_s\).
}
\label{fig:NuclearGeomScaling}
\end{center}
\end{figure}

Despite the relatively weak nuclear enhancement of saturation effects, it is important to point out, that the TMDs in our impact parameter dependent saturation model exhibit geometric scaling to a rather good accuracy.
This phenomenon can be observed in the plot in the bottom left panel of \reffig{NuclearGeomScaling} which shows the dipole TMDs as absolute values and as ratios to the proton case. Similar scaling behavior holds also for all other TMDs as shown in the top panels of Fig.~\ref{fig:NuclearGeomScaling}, where we present results for the seven TMDs relevant to forward di-jet production for various nuclei.
Importantly, in all of the plots the nucleus TMDs are rescaled along the \(x\) and \(y\) axes according to
\begin{gather}
F(x,k_t,A)\to \l.F\lr{x,\f{k_t}{Q^A_{s}(x)}}\middle/\rho_\perp^{A}\r.\;,\label{eq:model-rescaling}
\end{gather}
such that the result for the various TMDs in Fig.~\ref{fig:NuclearTMDSummary} nearly collapse onto a single curve for all the different nuclei. While this clearly indicates the property of geometric scaling, it is important to point out once again, that the nuclear enhancement of saturation effects is rather mild, and merely gives rise to about a factor of \(Q^{Pb}_s/Q^p_s\sim 2\) for the Lead, as can be seen in the bottom panel of Fig.~\ref{fig:NuclearGeomScaling}, where we show the $x$ dependence of the nuclear saturation scale $(Q_{s}^{A})^2(x)$.

\begin{table}[htbp]
  \begin{center}
    \begin{tabular}{|p{1.5cm}|p{1.5cm} p{1.5cm} p{1.5cm}  p{1.5cm} p{1.5cm} p{1.5cm}|}
      \hline
      N & \(\A^-\) & \(\A^=\) & \(\A^+\) &\(\T^-[\mathrm{fm}^{-2}]\) &\(\T^=[\mathrm{fm}^{-2}]\) &\(\T^+[\mathrm{fm}^{-2}]\)\\
      \hline
      \(p\) & 0.902 & 0.863 & 0.805 & 0.562 & 0.582  & 0.612 \\
      \(O\) &  14.3 & 14.1 & 13.7 & 0.975 & 0.987 & 1.01 \\
      \(Xe\) & 126 & 126 & 125 & 1.76 & 1.77 & 1.78 \\
      \(Pb\) & 205 & 204 & 203 & 1.99 & 1.99 & 1.99 \\
      \hline
    \end{tabular}
    \caption{Nuclear density normalization \(\A^{-=+}_A=\A_A(T^{-=+}_{\rm cut})\) and average thickness value \(\T^{-=+}_A=\T_A(T^{-=+}_{\rm cut})\) where the superscripts \(-,=,+\) indicate the lower, central and upper \(T_{\rm cut}\) values respectively.}
  \label{tab:rescaling}
  \end{center}
\end{table}

\section{Di-jet production at the LHC}
\label{sec:dilute-dense}

This section is dedicated to study how the model predictions for the impact parameter dependent TMDs, detailed in the previous section, may affect the search for saturation effects at hadron colliders.
To that end, we first introduce the selected experimental observables and then we discuss the results with emphasis on the distinctive features that separate the model predictions from what a simpler naive scaling approximation would have suggested.

The forward di-jet production in proton-proton and proton-nucleus collisions is one of the preferred processes\cite{Gribov83:QCDSemihardProcesses} through which the target (proton or nucleus) can be probed at very small longitudinal fractions, i.e.\ when the gluon density is large and recombination mechanisms related to saturation are expected to appear especially in heavy nuclei.
We consider the special kinematics in hadronic ($p-p$ or $p-A$) collisions when both  partons (jets) are measured at large rapidities in the same forward hemisphere.
The strongly asymmetric configuration in jet emission results into a \emph{dilute-dense} asymmetry in the partonic distributions of target and probe that interact in the scattering, which is due to the strong imbalance between the
longitudinal momentum factions (\(x_2\ll x_1\)) of the respective interacting partons.

At high energies, the longitudinal momentum of the jets is entirely provided by the co-moving parent parton carrying sizable part of the proton momentum\footnote{The case with a large longitudinal momentum  fraction extracted from the parton moving in the opposite longitudinal direction is suppressed by a factor of \(1/s\).} which can, thus, be described in terms of the well known PDFs. On the other hand, the partonic content in the target (the counter-moving hadron) is probed at much smaller \(x\) values where effects of gluon saturation can be probed.

Besides for the longitudinal fractions \(x\) also the initial transverse momentum of the scattered gluon in the target \(k_t\) plays a crucial role in the sensitivity to saturation effects.
Its value can discriminate between different physical regimes that require descriptions withing different frameworks. On one hand, when it is of the order of the jet transverse momenta \(k_t\simeq p_j\) the usual high energy factorization applies~\cite{KutakSapeta12:LHC-GluonSaturation-pPb}. On the other hand, when \(k_t\) approaches the saturation scale \(Q_s\) in the target for which \(p_{j}\gg Q_s\sim k_t\), the process can be described in the framework of Transverse Momentum Dependent gluon distributions (TMDs) supplied with small-\(x\) corrections.

Quite recently, the Improved-TMD (ITMD) factorization scheme --- a convenient unifying framework that encompasses the two regimes and provides a smooth interpolation between the two descriptions --- was proposed in Ref.~\cite{KotkoKutakMarquet15:DiluteDenseImprovedTMDs}.
In this scheme, cross sections in the dilute-dense regimes are given as convolution of the probe (proton) PDF, the target (proton or nucleus) TMDs and several hard factors regardless of the relative size of \(k_t\) and \(p_j\) (provided that \(p_j\gg Q_s\)). Each TMD couples with a specific hard factor which represents the scattering off-shell matrix element. The universality of the partonic distribution on the target side is restored at large \(k_t\) where all the TMDs collapse to the same perturbative power law behavior, which appears in the high-energy factorization framework\footnote{Note that the functional form of the dipole distribution \(\N\) in the MVgamma model with the logarithmic drop for large dipole sizes \(\vec{r}\) is crucial to enable the matching of the TMDs in this limit.}
In Ref.~\cite{Marquet16:SmallxITMDs} this framework was applied to describe forward di-jet production in proton-proton (p-p) and proton-lead (p-Pb) collisions at center of mass energies \(\sqrt{s_{NN}}=8.16\) TeV at the LHC.

We follow this path and proceed to employ the ITMD framework in our analysis. The cross section (\refeq{cross-section}) and the hard factor (App.~\ref{sec:tmd-formulae}) formulae that we use are identical to those of Ref.~\cite{Marquet16:SmallxITMDs}. However,
in contrast to previous works\footnote{Besides for the nuclear mass dependence, our analysis differs from theirs also in the way the small-\(x\) evolution was accounted for.
Instead of the AAMQS fit based on the rcBK solution of the evolution equation they opted for a description of the evolution in terms of the KS gluon distribution which is solution to a different extension of the BK equation\cite{KutakSapeta12:LHC-GluonSaturation-pPb}.}, we employ the TMDs calculated in Sec.~\ref{sec:tmds} and App.~\ref{sec:tmd-calculation} within our impact parameter dependent description of the nucleus, rather than assuming ``naive'' scaling for the saturation scale $Q_{s}^2 \sim A^{1/3}$ and transverse area $S_{\bot} \sim A^{2/3}$. Consequently, important differences can be expected when considering the cross-section in \(p-A\) collisions.

\subsection{Determination of the forward jet production cross section}
\label{sec:cross-section}

\begin{figure}[ht]
  \begin{center}
    \includegraphics[width=0.7\textwidth]{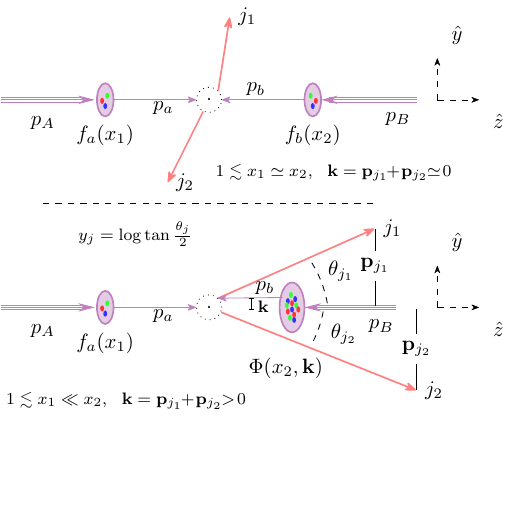}
      \vspace{-60pt}
    \caption{Schematic of central and forward di-jet hadro-production. Only the forward jet configuration can be sensitive to the saturation region \(x\ll 1, k_t\sim Q_s\).}\label{fig:geometry}
  \end{center}
\end{figure}

We will compute the cross section of inclusive di-jet production in proton-hadron collisions when both jets are emitted in the forward region~\footnote{With \(p\), we indicate a 3-momentum and we often parametrize it in terms of rapidity and transverse component with respect to the beam as \(p=(y,\vec{p})\).}
\[p(p_1)+A(p_2)\to j_1(p_{j_1})+j_2(p_{j_2})+X\;,\]
where \(X\) stands for anything (or possibly nothing) produced in addition to the two jets. In this process,
a projectile --- a proton with momentum \(p_1\) --- scatters against a target --- another proton (\(A=p\)) or a nucleus (\(A=\{O,~Xe,~Pb\}\)) with momentum \(p_2\) --- to produce two jets in the forward direction of the projectile (\(y_{j_1},y_{j_2}>0\)). \reffig{geometry} displays a schematic of the considered process.
We note that both jets are required to be ``hard'', i.e.  \(\abs{\vec{p}_j}\gg \Lambda_{\rm QCD}\).
Within our analysis, we will consider two sets of generic kinematic ranges within the acceptances of a typical detector at the LHC such as the  CMS or ATLAS central detector (CTR) or the CMS CASTOR~\cite{CMS20:CastorCalorimeter} and the ALICE FOCAL~\cite{Snowmass22:LHCForwardCalorimeter} (FWD) detectors (that allow to reach lower values of $x$) at a typical  center-of-mass energy of \(\sqrt{s}=8.16\) TeV. These are shown in Table~\ref{tab:acceptance}. Experimental data are already available for a forward jet measurement using the CMS CASTOR detector.
The acceptance of the forward detector in rapidity and transverse energy makes it the ideal settings to probe the lowest \(x\) tail of the target nuclear distribution.

The longitudinal momentum fraction (with respect to the parent) of the interacting parton in the projectile \(x_1\) and in the target \(x_2\) are defined as
\begin{subequations}
\begin{align}
x_1 & %=\frac{p_1^+ + p_2^+}{p_p^+}
=\frac{1}{\sqrt{s}} \left(\abs{\vec{p}_{j_1}} e^{y_{j_1}}+\abs{\vec{p}_{j_1}} e^{y_{j_2}}\right)\,, \\
x_2 & %= \frac{p_1^- + p_2^-}{p_A^-}
=\frac{1}{\sqrt{s}} \left(\abs{\vec{p}_{j_1}} e^{-y_{j_1}}+\abs{\vec{p}_{j_1}} e^{-y_{j_2}}\right)\,,\eq{x2}
\end{align}
\end{subequations}
where \(y_{j_1/2}\) and $\vec{p}_{j_{1/2}}$ are the jet rapidities and transverse momenta respectively.
With the acceptances given in Table~\ref{tab:acceptance}, the lowest possible  values of \(x_2\) allow probing the kinematical domain where saturation effects can appear
\begin{subequations}
  \begin{align}
    10^{-5}\lesssim &x_2 \lesssim 10^{-3}, & {\rm CTR}\\
    10^{-6}\lesssim &x_2 \lesssim 10^{-4}, & {\rm FWD}
  \end{align}
\end{subequations}
\begin{figure}[ht]
  \begin{center}
    \includegraphics[width=0.4\textwidth]{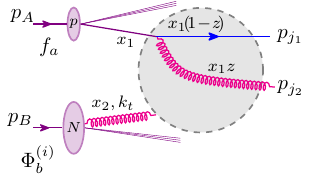}
    \caption{Schematic of the forward jet process In the TMD formalism, the interacting gluon in the target can be connected to the upper parton from the projectile as a gluon-parton scattering everywhere within the gray area.}\label{fig:cross-section}
  \end{center}
\end{figure}

The ``master-formula'' for the calculation of the cross section in the I-TMD factorization framework reads
\begin{equation}
\eq{cross-section}
% \frac{\dd\sigma^{pA\rightarrow {\rm di-jets}+X}}{\dd^{2}\vec{P}\dd^{2}\vec{k}\dd y_{j_1}\dd y_{j_2}}=
\frac{\dd\sigma^{pA\rightarrow {\rm di-jets}+X}}{\dd^{3}J_1\dd^{3}J_2}=
% \frac{\dd\sigma^{pA\rightarrow {\rm di-jets}+X}}{\dd^{2}\vec{p}_{j_1}\dd^{2}\vec{p}_{j_1}\dd y_{j_1}\dd y_{j_2}}=
\frac{\alpha_{s}^{2}}{(x_1 x_2 s)^{2}}
\sum_{a,c,d} \frac{x_1 f_{a/p}(x_1)}{1+\delta_{cd}}\sum_{i=1}^{2}K_{ag\to cd}^{(i)}(P_t,k_t)\Phi^{(i)}_{ag\rightarrow cd}(x_2,k_t)\;,
\end{equation}
where \(J=(\vec{p}_j,y_j),\dd^3J=\dd^2\vec{p}_j\dd y_j\) is a compact notation to indicate the jet kinematics. \(P_t=\abs{\vec{P}}\) is the jet hard scale given in \refeq{zdef}. \(k_t=\abs{\vec{k}}\) is the jet transverse momentum imbalance \(\vec{k}=\vec{p}_{j_1}+\vec{p}_{j_2}\) that can be traced back to the transverse momentum of the small $x$ gluon inside the target (as the other partons are considered to be collinear to the beam in this approximation)~\footnote{Strictly speaking, the correspondence between the jet imbalance and the initial parton momentum works only in absence of additional radiation or when the system recoil due to inclusive emissions \(X\) is negligible.}. Also, we impose $p_{j_1}>p_{j_2}$ in the phase-space integrals.
The \(\Phi^{(i)}_{ag\to cd}(k_t,x_2),\,i=1,2\,\) are special linear combination of the target TMDs that couple with the hard factors \(K^{(i)}_{ag^*\to cd}\) computed in the ITMD formalism. Their explicit expressions are given in Table~\ref{tab:Khardfactors} of Appendix~\ref{sec:tmd-formulae}.
The large \(x\) partons in the projectile are described by the usual collinear proton PDFs \(f_a(x_1,\mu_F)\). For the collinear PDFs, we use the CTEQ18\cite{Hou19:CteqGlobalAnalysis} parametrization. The number of active flavors is set to \(N_f=3\). We exclude the heavy flavors to match that setting in the TMD calculation (see. Sec.~\ref{sec:dipole-picture}). The same CTEQ18 package is used to compute the strong  coupling \(\alpha_s\).
The factorization and renormalization scales are defined as \(\mu_R=\mu_F=(p_{j_1}+p_{j_2})/2\).
Note that hadronization effects and jet recombination mechanisms are entirely neglected in this analysis.
In this approximation, a one-to-one correspondence between a given jet and its partonic parent can be drawn.
\begin{table}
  \begin{center}
    \begin{tabular}{ |p{2cm}|p{2cm}p{2cm}p{2cm}p{2cm}|  }
      \hline
      \multicolumn{5}{|c|}{Kinematic Ranges} \\
      % \hline
      &$y_{\rm Min}$&$y_{\rm Max}$&$p^{\rm Min}_{t}$ [GeV]  &$p^{\rm Max}_{t}$ [GeV]\\
      \hline
      CTR     &    \(3.5\)    &    \(4.5\)     & \(\{10, 20, 40\}\)    & \(\{20, 40, 80\}\) \\
      FWD &    \(5.2\)    &    \(6.6\)     & \(\{5, 10\}\)  & \(\{10,20\}\) \\
      \hline
    \end{tabular}
    \caption{Kinematic cuts corresponding to the acceptances of the CMS central detector (CTR) and ALICE/FOCAL and CMS/CASTOR forward detectors (FWD).}\label{tab:acceptance}
  \end{center}
\end{table}

\subsection{Results}
\label{sec:results}

In this section, we present the model predictions for the forward di-jet production in \(p-p\) and \(p-A\) collisions at \(\sqrt{s}={8.16}\) TeV. The goal is to observe the effects on the cross section of the dynamics of emission and recombination shaping the nuclear distribution in the target, thus probing the onset of the saturation regime.
A simple observable that is less sensitive to the precise measurement of the jet transverse momentum, difficult in the very forward region,
is the jet azimuthal decorrelation
\be
\eq{int-cross-section}
\f{\dd\sigma}{\dd\Delta\phi_{j_{1,2}}}=\intd^3J_1\dd^3J_2\,\delta\lr{\Delta\phi_{j_{1,2}}-(\phi_{j_1}-\phi_{j_2})}
\f{\dd\sigma}{\dd^{3}J_1\dd^{3}J_2}\;,
\ee
where the right hand side of this equation was given in \refeq{cross-section}.
Another observable is the nuclear modification factor
\be
\eq{RN/Ap}
R_{pA}=
% R_{\nicefrac{N}{Ap}}=
\l.\f{\dd\sigma^A}{\dd\Delta\phi_{j_{1,2}}}\middle/\!A\f{\dd\sigma^p}{\dd\Delta\phi_{j_{1,2}}}\r.\;,
\ee
which weights the saturation effects (normalized by number of nucleons) in nuclei versus proton targets. The larger saturation scale in nuclei is expected to result in a suppression towards \(\Delta\phi_{j_{1,2}}\sim\pi\), where the transverse momentum imbalance \(k_t\) can be on the order of the saturation energy \(k_t\sim Q_s\).

\begin{figure}[htbp]
  \begin{center}
    \includegraphics[width=0.8\textwidth]{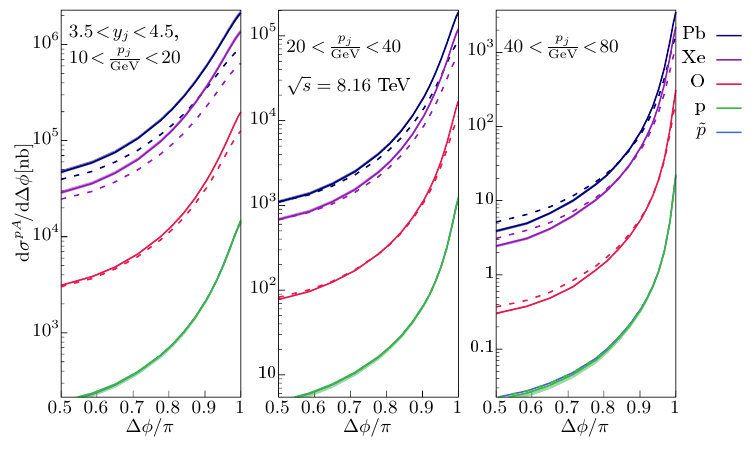}
    \caption{Azimuthal angle distributions for $p$, $\tilde{p}$, $Xe$, $O$ and $Pb$ for three bins in jet $p_j$ in the acceptance of the central detectors as predicted by the model  (solid bands) and in the ``naive'' scaling approximation (dashed lines). The \(\tilde{p}\) (no \(b\)-dependence) lays within the (\(b\)-dependent) \(p\) error band.}
\label{fig:cms-pA}
  \end{center}
\end{figure}

\begin{figure}[htbp]
  \begin{center}
    \includegraphics[width=0.7\textwidth]{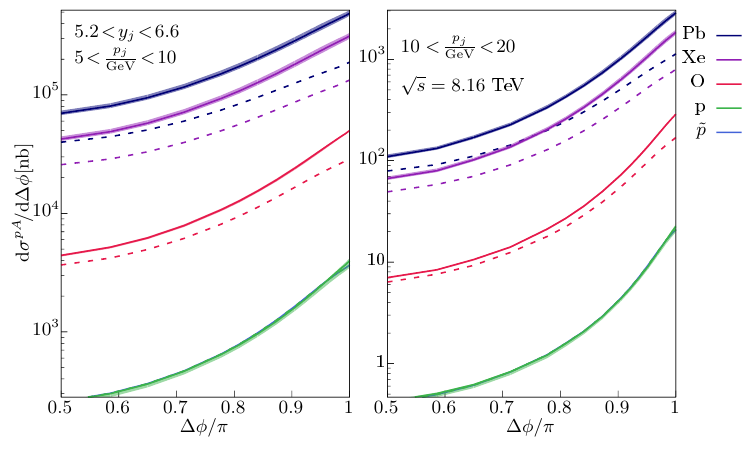}
    \caption{Azimuthal angle distributions for $p$, $\tilde{p}$, $Xe$, $O$ and $Pb$ for two bins in jet $p_j$ in the acceptance of the forward detectors  as predicted by the model  (solid bands) and in the ``naive'' scaling approximation (dashed lines).}\label{fig:castor-pA}
  \end{center}
\end{figure}

Differential cross sections as a function of the azimuthal angle between the two jets defined in \refeq{int-cross-section} for \(p-p\), \(p-O\), \(p-Xe\) and \(p-Pb\) are shown in Figs.~\ref{fig:cms-pA} and~\ref{fig:castor-pA} in the CTR and FWD detector kinematics respectively (see Table~\ref{tab:acceptance}). Different solid curves in Figs.~\ref{fig:cms-pA} and~\ref{fig:castor-pA}, show the results obtained using our impact parameter dependent model TMDs, whereas dashed lines show the results for the same nuclei, obtained using in \refeq{int-cross-section} the average proton TMDs rescalend as
\begin{equation}
F^{A}_{\rm naive}(k_t/Q_s(x),x)=A^{2/3}F^{\tilde{p}}\l(k_t\middle/\l[Q_s(x)A^{1/6}\r],x\r)\;. \eq{naive-tmds}
\end{equation}
In the case of the impact parameter dependent model, the colored bands in Figs.~\ref{fig:cms-pA} and~\ref{fig:castor-pA} indicate the uncertainties of the cross section due to the variation of the large distance cut-off \(T_{\rm cut}\) (see Sec.~\ref{sec:paramter-fix}).

Before we comment on the comparison between the two, we first explain the general features of the cross-sections in Figs.~\ref{fig:cms-pA} and~\ref{fig:castor-pA}. Indeed, the interpretation of the observed distribution behavior in the context of saturation dynamics is straightforward if one keeps in mind the connection between the investigated jet configurations and the two variables \(x_2\) and \(k_t\) that describe the target gluon distributions.
Firstly, the typical value of \(x\) at which the gluons in the target are resolved in the scattering depends on the jet rapidity and momentum ranges (see \refeq{x2}) and dictates the magnitude of the saturation scale $Q_{s}$.
In particular, looking at Figs.~\ref{fig:cms-pA} and~\ref{fig:castor-pA}, in this order, and at the plots within each figure from right to left we observe a decrease in the typical \(x\) value and, thus, an increase in \(Q_s(x)\) value at which the target is probed.
Secondly, looking at these plots in the same sequential order, we expect that, due to decreasing jet energies, the azimuthal region sensitive to saturation --- where the jet momentum imbalance is of the order of the saturation scale \(k_t\sim Q_s\) --- to slowly expand from the vicinity of the back-to-back region \(\Delta\phi_{j_{1,2}}=\pi\) towards smaller angles.
The consequence of these two reinforcing effects are the observed flatter distributions with shallower peaks at \(\Delta\phi_{j_{1,2}}=\pi\) when more forward and softer jets are measured.  Similarly, the proton-proton cross sections are more peaked around the back-to-back configuration compared to the proton-nucleus case, as expected since the saturation scale is smaller for the proton as compared to the heavier ions.
Even though these general features of the cross section can be observed for both the nuclear model TMDs (solid) and the naive scaling ansatz (dashed), there are discernible differences in the predictions for the cross section in proton-nucleus collisions. While for the case of the proton, the results for average proton TMDs fall within the uncertainty band obtained using the \(b\)-dependent proton model TMDs, differences between the ``naive'' and impact parameter dependent description, regarding both the normalization and shape of the cross section increase with increasing atomic number from Oxygen to Xenon and Lead. Notably, the largest discrepancies arise for low momentum jets and in the forward rapidity region, where the di-jet production is sensitive to small $x$ in the nucleus, and thus to the dynamics around the nuclear saturation scale $Q_s$.
However, important unaccounted corrections are expected to substantially affect the corresponding predictions since the underlying hypothesis \(p_j\gg Q_s\), laying the foundation of the TMD description, is no longer valid. In this regime, a complete description within the full-complexity of the CGC framework would be  required to take these corrections into account, and we refer the interested reader to~\cite{Fujii:2020bkl} for recent progress in this direction.
Conversely, for high momentum jets that mostly probe the perturbative tail of the TMDs, the agreement between the impact parameter dependent model and the naive scaling formula is rather good. We will continue on the comparison between naive \emph{versus} model predictions further down after commenting on the model predictions for the nuclear modification factors.

\begin{figure}[htbp]
  \begin{center}
    \includegraphics[width=0.75\textwidth]{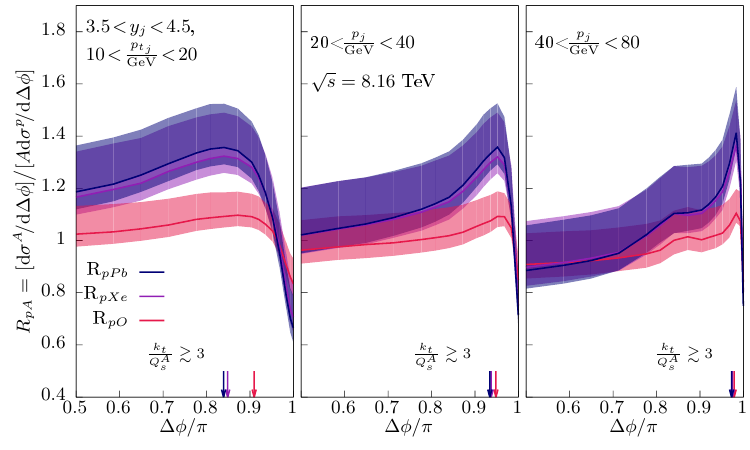}
    \caption{Oxygen, Xenon and Lead nuclear modification factors as a function of $\Delta \Phi$ between the jets for three bins in jet $p_j$ in the acceptance of the forward detectors. The colored bands span the model uncertainties. The colored arrows divide the azimuthal angular regions associated with the large tail regime and the saturation region.}\label{fig:cms-pp-vs-pA}
  \end{center}
\end{figure}

\begin{figure}[htbp]

  \begin{center}
    \includegraphics[width=0.7\textwidth]{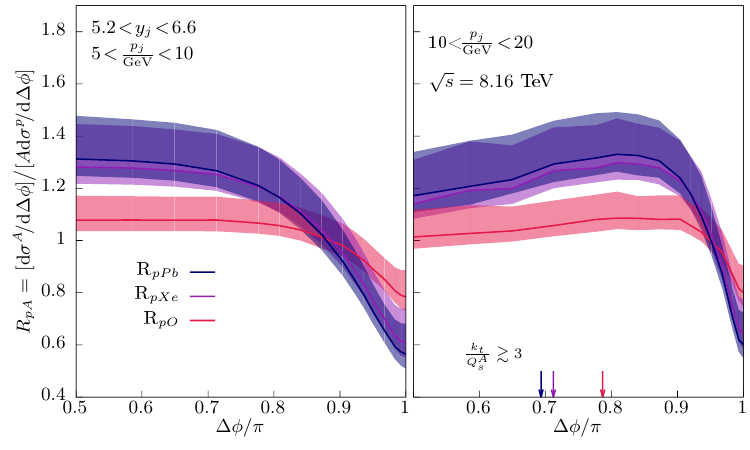}
    \caption{Oxygen, Xenon and Lead nuclear modification factors as a function of $\Delta \Phi$ between the jets for two bins in jet $p_j$ in the acceptance of the forward detectors. The colored bands span the model uncertainties. The colored arrows divide the azimuthal angular regions associated with the large tail regime and the saturation region.}\label{fig:castor-pp-vs-pA}
  \end{center}
\end{figure}

Evidently, the precise modeling of nuclear TMDs also has immediate consequences for the behavior of the nuclear modification factor $R_{pA}$. In Figs.~\ref{fig:cms-pp-vs-pA} and~\ref{fig:castor-pp-vs-pA} we show the nuclear modification factors for Oxygen, Xenon and Lead, in the two different kinematical ranges accessible within present and future LHC experiments. Once again, the colored bands represent the model uncertainties, associated with the \(\pm 50\%\) variation of the large distance cut-off in the nuclear TMD calculation~\footnote{
The upper edge (\(+\)) of the error bands is defined as  \[R^{+}_{pA}=\l.\f{\dd\sigma^A(T^{+}_{\rm cut})}{\dd\Delta\phi_{j_{12}}}\middle/A\f{\dd\sigma^p(T^{+}_{\rm cut})}{\dd\Delta\phi_{j_{12}}}\r.\,.\]
The lower edge (\(-\)) is defined likewise.}.
Interestingly, we find that the modification factors \(R_{pXe}\) and \(R_{pPb}\) are almost identical. On the other hand, \(R_{pO}\) remains around \(1\) within uncertainties except in the immediate vicinity of $\Delta\phi=\pi$ as it is best seen in \reffig{castor-pp-vs-pA}.  Generally, the $R_{pA}$ shows a modest suppression of back-to-back jets and enhancement of nearly back-to-back jets,  which can be attributed to the azimuthal de-correlation of the jet pair due to the transverse momentum $k_t$ provided by the nuclear small $x$ TMD gluon.
Since the typical transverse momentum $k_t$ is of the order of the saturation scale $Q_{s}(x)$, we use arrows (indicated by matching colors) to mark the azimuthal angle below which the respective nuclear TMDs are sampled at values \(k_t/Q^A_s\leq 3\) in the scattering.\footnote{To be more precise, the azimuthal angles marked by the arrows are found by solving the implicit equation \(\min(k_t/Q^A_s)=3\), where the left hand side is a function of   \(\Delta\phi_{j_{1,2}}\).} Naturally, we find that the location of the maxima coincides rather well with the position of the arrows, indicating that at the respective angular scale the measurement is indeed sensitive to saturation effects.

\begin{figure}[htbp]
  \begin{center}
    \includegraphics[width=0.75\textwidth]{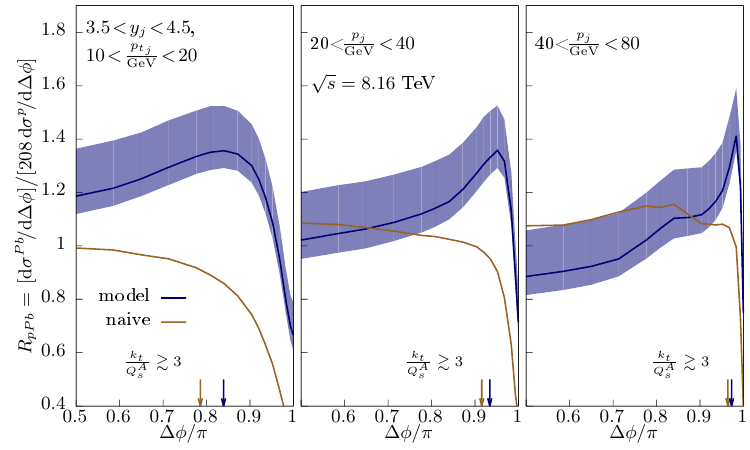}
    \caption{Nuclear modification factor \(R_{pPb}\) as a function of jet $\Delta \Phi$ in three bins in $p_j$ from our model compared to the ``naive'' scaling prediction in the acceptance of the central detectors. The vertical arrow gives a rough indication on the \(\Delta\phi^*\) value where saturation effects become important defined as  \(k_t/Q^A_s<3\) on the right from the arrow. It is not visible on the left plot where \(k_t/Q^A_s<3\) is everywhere.}\label{fig:cms-model-vs-naive}
  \end{center}
\end{figure}

\begin{figure}[htbp]
  \begin{center}
    \includegraphics[width=0.7\textwidth]{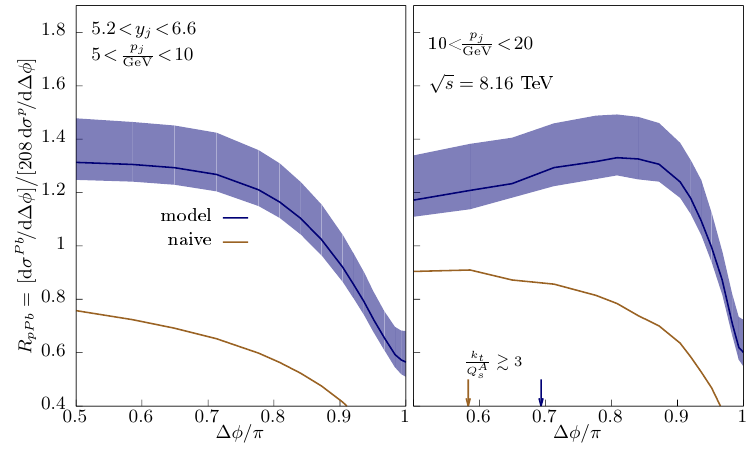}
    \caption{Nuclear modification factor \(R_{pPb}\) as a function of jet $\Delta \Phi$ in two bins in $p_j$ from our model compared to the ``naive'' scaling prediction in the acceptance of the forward detectors.
      The vertical arrow gives an  indication on the \(\Delta\phi^*\) value where saturation effects become important  defined as  \(k_t/Q^A_s<3\) on the right from the arrow.  It is not visible on the left plot where \(k_t/Q^A_s<3\) is everywhere.}\label{fig:castor-model-vs-naive}
  \end{center}
\end{figure}

Now that we have established the basic features of the nuclear modification of di-jet production, we will contrast our results from the impact parameter dependent TMD model with those obtained using the ``naive'' scaling approximation, where the nuclear TMDs are obtained by scaling the saturation scale of the nucleus $(Q^A_{s})^2=A^{1/3} (Q^{p}_{s})^2$. We exemplify this for the Lead nuclear modification factor shown in Figs.~\ref{fig:cms-model-vs-naive} and~\ref{fig:castor-model-vs-naive}, where the nuclear modification factor in the ``naive'' scaling approximation (brown line) is calculated as in \refeq{RN/Ap}
using the TMDs defined in \refeq{naive-tmds}.
By comparing the results, one observes that in the saturation region, i.e. at low momentum, forward rapidity or close to the back to back limit, the model predictions differ significantly from the ``naive'' scaling expectation. Conversely, for large momentum imbalance and at large jet momentum, where one probes the perturbative tail of the gluon distributions, the model predicition is in reasonable agreement with the naive scaling prediction. Clearly, the different behavior of the two models can be directly understood from the behavior of the underlying nuclear TMDs  in the top left panel of \reffig{NuclearTMDSummary}. 

While the perturbative tails match between the model and the naive scaling predicition, the impact parameter dependent model predicts broader and higher peaks of the nuclear TMDs (solid lines) around the saturation scale as compared to the naive scaling (dashed lines) expectations. Noteably, the larger ratio of the nuclear TMDs to proton TMDs, also shown in \reffig{NuclearTMDSummary}, is directly reflected in the ratio of the cross-sections $R_{pA}$, and gives rise  to  a ``knee'' shaped enhancement of the nuclear modification factor at intermediate angles, which is a distinctive feature of the model predictions that is not present with the naive scaling of the saturation scale of the nucleus. Generally, we find that the suppression ($R_{pA}<1$) close to back-to-back kinematics ($\Delta \phi \simeq pi$) appears to be a robust model independent feature, which can be attributed to the larger transverse momentum acquired from multiple scattering in the nucleus. However, our analysis also shows that the behaviour at intermediate angles can in fact be quite sensitive to the behaviour of nuclear TMDs, and can vary from enhancement ($R_{pA}>1$) obtained in the presumably more realistic impact parameter dependent model, to suppression ($R_{pA}<1$)  obtained with naive scaling of the saturation scale.

\section{Conclusions}
\label{sec:conclusions}

We performed a phenomenological analysis of forward dijet production in \(p-p\) and \(p-A\) (Oxygen, Xenon and Lead nuclei) collisions in dilute-dense kinematics which can be understood within the ITMD framework. The goal is to analyze the effects on the dynamics of saturation due to the complex transverse distribution of color charges in nuclei.

We developed a simple model to compute the TMDs for any nucleus given its nuclear matter distribution. By assuming Gaussian statistics for the nuclear color charge distributions, all the small-x TMDs can be computed from the knowledge of the fundamental nuclear dipole amplitude \(N(x,r)\), where \(r\) is the dipole size, which can be inferred with great precision and down to very small values of \(x\) from DIS measurements at HERA. One typical example, which we make use of, is the AAMQS fit using rcBK evolution equation.

Since saturation model fits to HERA data are often performed at the level of inclusive cross-section, and thus not sensitive to the impact parameter dependence of saturation effects, extending the calculations to the complex geometry of heavy and light nuclei is non-trivial and can result in a significant uncertainty of the predictions.  Inspired by the IPSat model, we therfore propose to re-introduce the impact parameter dependence at the level of the dipole amplitude, to consistently derive TMDs of heavy and light nuclei from the underlying proton distributions.
This procedure introduces a few free parameters that we omit here, which can be fixed by matching (the transverse integral of) the \(b\)-dependent TMDs with the \(b\) averaged TMDs computed from the inclusive dipole amplitude provided by the underlying AAMQS fit.
As cross-check, we verified that this procedure can still reproduce the proton structure function \(F_2\) well.
Clearly the advantage of the impact parameter dependent distributions is that they can be straightforwardly generalized to nuclei, and the respective TMDs can be easily obtained by finally integrating over the impact parameter. Based on this impact parameter dependent treatment of saturation effects, the newly found TMDs show a dependence upon the nuclear mass number that deviate significantly from a naive scaling expectation. More precisely, the dependence on the atomic number \(A\) is stronger then \(A^{\nf{2}{3}}\) for the overall magnitude of the TMD but it is weaker then \(A^{\nf{1}{6}}\) for the characteristic momentum \(Q^A_s\). These effects also have important consequences on the \(p-A\) scattering cross sections.

The forward di-jet cross section in \(p-p\) and \(p-A\) collisions are constructed convoluting (the impact parameter average of) the nuclear TMDs with the appropriate hard factors in the usual ITMD scheme. We presented predictions corresponding to the kinematics acceptance of the forward portion of the CMS detector and the Castor (CMS)/Focal (ALICE) very forward calorimeters.

As expected the $p-A$ and even more so $p-p$ cross sections are strongly peaked at $\Delta\phi_{j_{12}}=\pi$ corresponding to small gluon transverse momenta $k_t$ in the target. The peaks are more pronounced for the highest jet momentum and more central rapidity acceptances since in this case the azimuthal angular region sensitive to saturation is squeezed towards the back-to-back configuration.
Furthermore, we observe that
the \(p-A\) cross sections and the nuclear modification factors deviate significantly from the naive scaling approximation, especially when the jets are softer and more forward.
We confirm that, the earlier onset of saturation in nuclei versus protons induces a suppression on the nuclear modification factors \(R_{pA}\) towards \(\Delta\phi\sim \pi\).
However, the model predictions deviate significantly from what the naive scaling would suggest. A ``knee'' region develops while moving from small to large azimuthal angles, before the saturation regime is reached and the nuclear modification factors are suppressed. At intermediate angles \(R_{pA}\) grows and can exceed unity (for \(Xe\) and \(Pb\), less so for \(O\)) before starting to drop. This effect is due to the broader and higher peaks that the impact parameter dependent TMDs show with respect to naive expectations. We therefore conclude, that a proper treatment of the impact parameter dependence of the nuclear matter distribution inside nuclei is in fact rather important in order to provide accurate predictions for saturation effects in nuclei.

Beyond the inclusive predictions presented here, it is also important to stress, that a proper treatment of the impact parameter dependence is crucial to also provide predictions for forward-central correlations, where e.g. the de-correlation of forward di-jets in $p-A$ collisions can be investigated as a function of the central event activity to enhance saturation signals. Even though such calculations will require further modeling of the event activity in the central detector region, our calculations clearly provide a first important step in this direction.

We finally remark, that in addition to a proper treatment of the nuclear geometry, there are other important phenomena that were not take into account in our analysis but can be expected to affect the predictions appreciably. First of all, an important unaccounted contribution is the already mentioned correction due to Sudakov resummation of $\log(p_j/k_t)$ enhanced diagrams. This resummation can be obtained in parallel with the BK $\log(1/x)$ resummation\cite{MuellerXiaoYuan13:Small-xSudakovResummation}, and its effects in this context have recently been investigated in Refs.~\cite{Caucal:2022ulg, Al-Mashad:2022zbq}.
We also did not include the heavy flavor ($c$ and $b$) quarks in the present TMD calculations, and we intend to return to both of these points in a forthcoming work.

We acknowledge support by the Deutsche Forschungsgemeinschaft (DFG) under grant CRC-TR 211
“Strong-interaction matter under extreme conditions” project no. 315477589-TRR 211. We also thank Cyrille Marquet and Martin Hentschinski for useful discussions at the beginning of this study.

\appendix

\section{Nuclear Thickness in the MC Glauber Model}
\label{sec:nuclear-matter}
Below we explain how the nuclear thickness \(T_A\),which describes the nuclear matter density projected in the transverse impact parameter space, is determined within the MC Glauber Model~\cite{Miller07:ReviewGlauberModel}. Starting point is the distribution of nucleons in side a nucleus, which is generated stochastically, event-by-event, by a Glauber Monte-Carlo sampling where, for each nucleon, random values in the range \(-5R,5R\) are assigned to each of its 3-dimensional coordinates (\(\underline{x}=\{x,y,z\}\)) in a system of reference with its origin placed at the nucleus center of mass. Subsequently, the probability of having a nucleon at such \(\underline{x}\) position is weighted by the Wood-Saxon (WS) distribution
\begin{equation}
\rho_{\rm W.S.}(r=\sqrt{\underline{x}^2};R,a)
=\f{\rho_0}{1+\exp\lr{\f{r-R}{a}}}\;,
\end{equation}
where \(R\) is the nuclear radius and \(a\) is the ``surface diffusiveness'' which characterizes the steepness of the distribution drop out in the proximity of its edges. The normalization constant in the Monte Carlo sampling is given by
\(\rho_0=1+\exp(-R/a)\), and corresponds to the nucleon density in the center of the nucleus.
\begin{table}[htbp]
  \begin{center}
    \begin{tabular}{|p{1cm} |p{1cm} p{1cm} p{1cm}|}
    % \begin{tabular}{|p{1cm}||p{1cm}|p{1cm}|p{1cm}|}
      \multicolumn{4}{c}{Wood-Saxon Distribution Params.} \\
      \hline
      N & \(A\) & \(R[{\rm fm}]\) & \(a[{\rm fm}]\) \\
      \hline
      % \(p\) & 1 & 1.22 & 0.4 & 0.18\\
      % \(He\) & 4 &  & 1.98 & 0.5\\
      \(O\) & 16  & 2.61 & 0.51\\
      \(Cu\) & 63 & 4.20 & 0.60\\
      \(Xe\) & 129  & 5.42 & 0.57\\
      % \(Au\) & 197 &  & 6.38 & 0.54\\
      \(Pb\) & 208 & 6.62 & 0.55\\
      \hline
    \end{tabular}
    \caption{All nucleus radii \(R\) and surface diffusiveness \(a\) parameters (see ~\cite{Miller07:ReviewGlauberModel})}
  \label{tab:params}
  \end{center}
\end{table}
Numerical values of the Wood-Saxon parameters for all nuclei are collected in Tab.~\ref{tab:params}.

Based on the nucleon positions, the nuclear thickness \(T_A\), is then found by convoluting the nucleon distribution in nucleus with the transverse matter density within each nucleon, i.e.
 given the nucleon distribution at the \(n\)-th MC iteration
\[\rho^{(n)}_{\rm W.S.}\lr{\underline{x}_1,\dots,\underline{x}_A}\;,\]
where \(\underline{x}_i\) with \(i=1\dots A\) specify all the nucleon positions,
the nuclear thickness is given by the superposition of the corresponding proton (or neutron) distributions
\begin{equation}
  \label{eq:MCGluberNuclearThickness}
  T_A(\vec{b})=\sum^{A}_{i=1} T_{p/n}(\abs{\vec{b}_i-\vec{b}})\;,
\end{equation}
where \(\vec{b}_i\) are the nucleon positions \(underline{x}_i\) projected over the transverse space. We employ a Gaussian profile with a width of \(\sqrt{B_G}=4 {\rm GeV}^{-1}\). for the transverse profile $T_{p/n}$ of the nucleon, with no distinction between protons and neutrons (\(T_p=T_n\)), with
\[T_p(\abs{\vec{b}})=\f{e^{-\vec{b}^2/\lr{2B_G}}}{2\pi B_G}\;.\]

\section{Hard Factors}
\label{sec:tmd-formulae}

All the expression collected in this section were taken from Ref.~\cite{KotkoKutakMarquet15:DiluteDenseImprovedTMDs}. They are reported here for convenience of consultation.

The gluon TMDs \(\Phi^{(i)}\) are given by a linear combination of the actual TMDs \(F^{(i)}_{xx}\) introduced in Sec.~\ref{sec:tmd-calculation}. They are collected in Table~\ref{tab:Khardfactors} together with the corresponding hard factors \(K^{(i)}\) which are given in terms of the Maldestan invariants
\begin{gather}
  \begin{subequations}\label{eq:maldestan}
    \begin{align}
      \shat &=(\vec{p}+\vec{k})^2 =(\vec{p}_1+\vec{p}_2)^2=\f{|\vec{P}|^2}{z(1-z)}\,, \\
      \that &=(\vec{p}_2-\vec{p})^2=(\vec{p}_1-\vec{k})^2=-\frac{|\vec{p}_{j_2}|^2}{1-z}\,, \\
      \uhat &=(\vec{p}_1-\vec{p})^2=(\vec{p}_2-\vec{k})^2=-\frac{|\vec{p}_{j_1}|^2}{z}\,,
    \end{align}
  \end{subequations}
  \shortintertext{and}
  \begin{subequations}
    \begin{align}
      \tls &= (x_2 p_{\! A}+p)^2
             =\frac{\vec{P}^2}{z(1-z)}+\vec{k}^2=x_1x_2s\,,\\
      \tlt &= (x_2 p_{\! A}-p_1)^2=-z\tls\,, \\
      \tlu &= (x_2 p_{\! A}-p_2)^2=-(1-z)\tls\,,
    \end{align}
    \label{eq:gen-mandelstam}
  \end{subequations}
\shortintertext{with}
z=\frac{p_1^+}{p_1^+ + p_2^+} \quad\quad \text{and}
\quad\quad \vec{P}=(1-z)\vec{p}_{j_1}-z\vec{p}_{j_2}\;.
\label{eq:zdef}
\end{gather}
Eqs.~\ref{eq:maldestan} sum up to $\shat+\that+\uhat=k_t^2$ and \(\tls+\tlt+\tlu=0\).

\renewcommand{\arraystretch}{3}
\begin{table}[htbp]
%\begin{doublespace}
\begin{centering}
  % \begin{tabular}{ | c{2cm} | c{5cm} c{5cm}| }
\begin{tabular}{|c|c c|}
\hline
\Large $i$ & \Large 1 &  \Large 2  \\
\hline
$ \Phi_{gg\to gg}^{(i)}$ &
\(\begin{aligned}
&\frac{1}{2N^2_c}\Big[N^2_c\left(\mathcal{F}_{gg}^{(1)}+\mathcal{F}_{gg}^{(6)}\right) \\
  &\qquad-2\mathcal{F}_{gg}^{(3)}+\mathcal{F}_{gg}^{(4)}+\mathcal{F}_{gg}^{(5)}\Big]
\end{aligned}\)
& \(\begin{aligned} &\frac{1}{N^2_c}\Big[N^2_c\left(\mathcal{F}_{gg}^{(2)}+\mathcal{F}_{gg}^{(6)}\right) \\
&\qquad -2\mathcal{F}_{gg}^{(3)}+\mathcal{F}_{gg}^{(4)}+\mathcal{F}_{gg}^{(5)}\Big] \end{aligned}\) \\[3ex]
\hline
$ \Phi_{gg\to q\bar{q}}^{(i)}$ & $ \frac{1}{N^2_c-1}\left(N^2_c\mathcal{F}_{gg}^{(1)}-\mathcal{F}_{gg}^{(3)}\right)$ & $ -N^2_c\mathcal{F}_{gg}^{(2)}+\mathcal{F}_{gg}^{(3)}$
\\
\hline
$ \Phi_{qg\to qg}^{(i)}$ & $  \mathcal{F}_{qg}^{(1)}$ & $ \frac{1}{N^2_c-1}\left(N^2_c\mathcal{F}_{qg}^{(2)}-\mathcal{F}_{qg}^{(1)}\right)$
\\
\hline
\hline
$ K_{gg^*\to gg}^{(i)}$ & $
\frac{N_{c}}{C_F}\,\frac{\left(\bar{s}^{4}+\bar{t}^{4}+\bar{u}^{4}\right)\left(\bar{u}\hat{u}+\bar{t}\hat{t}\right)}{\tlt\that\tlu\uhat\tls\shat}$ & $ -\frac{N_{c}}{2C_F}\,\frac{\left(\bar{s}^{4}+\bar{t}^{4}+\bar{u}^{4}\right)\left(\bar{u}\hat{u}+\bar{t}\hat{t}-\bar{s}\hat{s}\right)}{\tlt\that\tlu\uhat\tls\shat}$
\\
\hline
$ K_{gg^*\to q\bar{q}}^{(i)}$ & $ \frac{1}{2N_{c}}\,\frac{\left(\bar{t}^{2}+\bar{u}^{2}\right)\left(\bar{u}\hat{u}+\bar{t}\hat{t}\right)}{\bar{s}\hat{s}\hat{t}\hat{u}}$ & $ \frac{1}{4N_{c}^2 C_F}\,\frac{\left(\bar{t}^{2}+\bar{u}^{2}\right)\left(\bar{u}\hat{u}+\bar{t}\hat{t}-\bar{s}\hat{s}\right)}{\bar{s}\hat{s}\hat{t}\hat{u}}$
\\
\hline
$ K_{qg^*\to qg}^{(i)}$ & $ -\frac{\bar{u}\left(\bar{s}^{2}+\bar{u}^{2}\right)}{2\bar{t}\hat{t}\hat{s}}\left(1+\frac{\bar{s}\hat{s}-\bar{t}\hat{t}}{N_{c}^{2}\ \bar{u}\hat{u}}\right)$ & $ -\frac{C_F}{N_c}\,\frac{\bar{s}\left(\bar{s}^{2}+\bar{u}^{2}\right)}{\bar{t}\hat{t}\hat{u}}$
\\
\hline
\end{tabular}
\par\end{centering}
\caption{Hard factors \(K_{ag^*\to cd}^{(i)}\) and the gluon TMDs $\Phi_{ag\to cd}^{\left(i\right)}$ in the finite-$N_c$ limit.}
\label{tab:Khardfactors}
\end{table}

\section{TMD calculation}
\label{sec:tmd-calculation}

In this appendix we describe the calculation techniques that have been employed in order to compute the gluon TMDs needed for the analysis which were briefly introduced in Sec.~\ref{sec:tmds}.

Starting point of our discussion is the operator definition of the TMDs in the small-x limit, where the relevant TMDs for di-jet production take the form Ref.~\cite{MarquetPetreskaRoiesnel16:TMDsFromJIMWLK}
\begin{align}
\frac{g^2 (2\pi)^3}{4 S_\bot} F^{(1)}_{qg}(\kt)&= \qquad \intd^2(\xt-\yt)~e^{-i\kt(\xt-\yt)}~\tr\Big[ \left(\partial_{i}^{\xt} \Vd_{\xt}\right) \left(\partial_{i}^{\yt}  \V_{\yt}\right) \Big]\;, \\
\frac{g^2 (2\pi)^3}{4 S_\bot} F^{(2)}_{qg}(\kt)&=\frac{-1}{N_c} ~\intd^2(\xt-\yt)~e^{-i\kt(\xt-\yt)}~\tr\Big[ \left(\partial_{i}^{\xt} \V_{\xt}\right) \Vd_{\yt} \left(\partial_{i}^{\yt}  \V_{\yt}\right) \Vd_{\xt} \Big] \tr\Big[\V_{\yt}\Vd_{\xt}\Big]\;,
% \end{align}
\intertext{for the $qg$ channel and for the $gg$ channel}
% \begin{align}
\frac{g^2 (2\pi)^3}{4 S_\bot} F^{(1)}_{gg}(\kt)&=\frac{+1}{N_c}~ \intd^2(\xt-\yt)~e^{-i\kt(\xt-\yt)}~\tr\Big[ \left(\partial_{i}^{\xt} \Vd_{\xt}\right) \left(\partial_{i}^{\yt}  \V_{\yt}\right) \Big]\tr\Big[\V_{\xt}\Vd_{\yt}\Big]\;, \\
\frac{g^2 (2\pi)^3}{4 S_\bot} F^{(2)}_{gg}(\kt)&=\frac{-1}{N_c}~ \intd^2(\xt-\yt)~e^{-i\kt(\xt-\yt)}~\tr\Big[ \left(\partial_{i}^{\xt} \V_{\xt}\right) \Vd_{\yt} \Big]\tr\Big[\left(\partial_{i}^{\yt}  \V_{\yt}\right) \Vd_{\xt}\Big]\;, \\
\frac{g^2 (2\pi)^3}{4 S_\bot} F^{(3)}_{gg}(\kt)&=-\quad \intd^2(\xt-\yt)~e^{-i\kt(\xt-\yt)}~\tr\Big[ \left(\partial_{i}^{\xt} \V_{\xt}\right) \Vd_{\yt} \left(\partial_{i}^{\yt}  \V_{\yt}\right) \Vd_{\xt} \Big]\;, \\
\frac{g^2 (2\pi)^3}{4 S_\bot} F^{(4)}_{gg}(\kt)&=-\quad  \intd^2(\xt-\yt)~e^{-i\kt(\xt-\yt)}~\tr\Big[ \left(\partial_{i}^{\xt} \V_{\xt}\right) \Vd_{\xt} \left(\partial_{i}^{\yt}  \V_{\yt}\right) \Vd_{\yt} \Big]\;, \\
\frac{g^2 (2\pi)^3}{4 S_\bot} F^{(5)}_{gg}(\kt)&=-\quad  \intd^2(\xt-\yt)~e^{-i\kt(\xt-\yt)}~\tr\Big[ \left(\partial_{i}^{\xt} \V_{\xt}\right) \Vd_{\yt} \V_{\xt}\Vd_{\yt} \left(\partial_{i}^{\yt}  \V_{\yt}\right) \Vd_{\xt} \V_{\yt} \Vd_{\xt} \Big]\;, \\
\frac{g^2 (2\pi)^3}{4 S_\bot} F^{(6)}_{gg}(\kt)&=\frac{-1}{N_c^2} \intd^2(\xt-\yt)~e^{-i\kt(\xt-\yt)}~\tr\Big[ \left(\partial_{i}^{\xt} \V_{\xt}\right) \Vd_{\yt} \left(\partial_{i}^{\yt}  \V_{\yt}\right) \Vd_{\xt} \Big] \tr\Big[ \V_{\xt}\Vd_{\yt} \Big] \tr\Big[ \V_{\yt} \Vd_{\xt} \Big]\;,
\end{align}
where $S_\bot$ denotes the transverse area of the target.

\begin{figure}[t!]
\begin{center}
\includegraphics{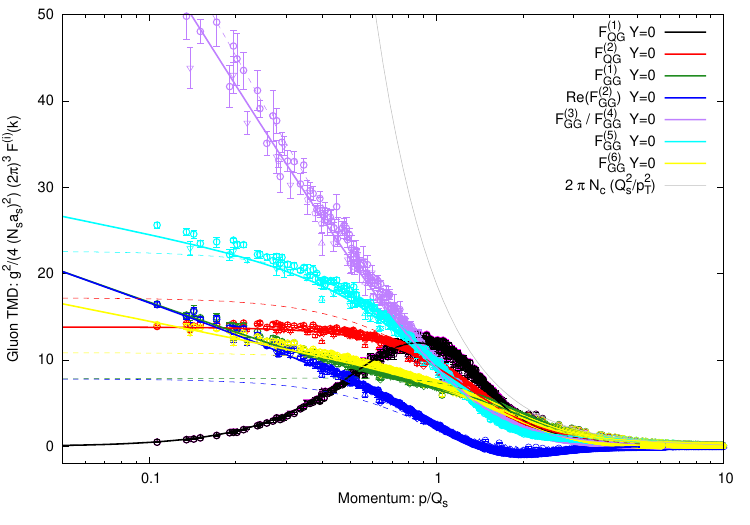}
\caption{\label{fig:TMDs_MV} Gluon TMDs in the MV model. Data points correspond to numerical lattice simulation. Solid (dashed) lines show the result obtained from the Gaussian appproximation at finite $N_c$ (in the large $N_c$ limit).}
\end{center}
\end{figure}

We have calculated the TMDs using an ensemble of lattice Wilson lines (see Sec.~\ref{sec:lattice-tmd-calc}) as well as in the Gaussian approximation (see Sec.~\ref{sec:gaussian-tmd-calc}). Note that in the large $N_c$ limit, expressions for the various TMDs have been provided in the literature, and the structure is simple especially in coordinate space (Eqs.~\ref{eq:largeNc-tmds-qg},\ref{eq:largeNc-tmds-gg}). In the $qg$ channel, they are given by
\begin{gather}
\begin{subequations}\label{eq:largeNc-tmds-qg}
\begin{align}
\frac{1}{N_c} \tilde{F}^{(1)}_{qg}(\xt,\yt)=&e^{G_{\xt\yt}} \left(G^{(i,i)}_{\xt\yt}-G^{(i,0)}_{\xt\yt} G^{(i,0)}_{\xt\yt}\right)\;, \\
\frac{1}{N_c} \tilde{F}^{(2)}_{gq}(\xt,\yt)=&\frac{e^{G_{\xt\yt}} \left(e^{2 G_{\xt\yt}}-1\right) G^{(i,i)}_{\xt\yt}}{2 G_{\xt\yt}}\;,
\end{align}
\end{subequations}
\intertext{while, in the $gg$ channel, they are}
\begin{subequations} \label{eq:largeNc-tmds-gg}
\begin{align}
\frac{1}{N_c} \tilde{F}^{(1)}_{gg}(\xt,\yt)=&e^{2 G_{\xt\yt}} \left(G^{(i,i)}_{\xt\yt}-G^{(i,0)}_{\xt\yt} G^{(i,0)}_{\xt\yt}\right)\;, \\
\frac{1}{N_c} \tilde{F}^{(2)}_{gg}(\xt,\yt)=&e^{2 G_{\xt\yt}} G^{(i,0)}_{\xt\yt} G^{(i,0)}_{\xt\yt}\;, \\
\frac{1}{N_c} \tilde{F}^{(3)}_{gg}(\xt,\yt)=&\frac{\left(e^{2 G_{\xt\yt}}-1\right) G^{(i,i)}_{\xt\yt}}{2 G_{\xt\yt}}\;, \\
\frac{1}{N_c} \tilde{F}^{(4)}_{gg}(\xt,\yt)=&\frac{\left(e^{2 G_{\xt\yt}}-1\right) G^{(i,i)}_{\xt\yt}}{2 G_{\xt\yt}}\;, \\
\frac{1}{N_c} \tilde{F}^{(5)}_{gg}(\xt,\yt)=&\frac{e^{2 G_{\xt\yt}} \left(4 e^{2 G_{\xt\yt}} G^{(i,i)}_{\xt\yt} G_{\xt\yt}^2+4 G^{(i,0)}_{\xt\yt} G^{(i,0)}_{\xt\yt} G_{\xt\yt}+\left(e^{2 G_{\xt\yt}}-1\right) G^{(i,i)}_{\xt\yt}\right)}{2G_{\xt\yt}}\;, \\
\frac{1}{N_c} \tilde{F}^{(6)}_{gg}(\xt,\yt)=&\frac{e^{2 G_{\xt\yt}} \left(e^{2 G_{\xt\yt}}-1\right) G^{(i,i)}_{\xt\yt}}{2 G_{\xt\yt}} \;.
\end{align}
\end{subequations}
\end{gather}
We also provide explicit coordinate space expressions for the various TMDs at finite $N_c$ (Eqs.~\ref{eq:finiteNc-tmds-qg},\ref{eq:finiteNc-tmds-gg}), which are significantly more complicated. Based on the calculation described in Sec.~\ref{sec:gaussian-tmd-calc}, one finds for the $qg$ channel
\begin{gather}
\begin{subequations}\label{eq:finiteNc-tmds-qg}
\begin{align}
\frac{1}{N_c} \tilde{F}^{(1)}_{qg}(\xt,\yt) &=\Big(  G^{(i,i)}_{\xt\yt} -  G^{(i,0)}_{\xt\yt} G^{(i,0)}_{\xt\yt}\Big)  e^{G_{\xt\yt}}\;, \\
  \begin{split}
    \frac{1}{N_c} \tilde{F}^{(2)}_{gq}(\xt,\yt)&=
    \left( (N_c+2)(N_c-1)e^{\frac{3N_c-1}{N_c^2-1} G_{\xt\yt}} + (N_c-2)(N_c+1)^2 e^{\frac{3N_c+1}{N_c^2-1} G_{\xt\yt}} -2N_c(N_c^2-3)e^{G_{\xt\yt}}\right)\\
    &\qquad\times\f{G^{(i,i)}_{\xt\yt}}{4N_c(N_c^2-1)G_{\xt\yt}}
    + \frac{1}{N_c^2-1} \left( G^{(i,i)}_{\xt\yt} - G^{(i,0)}_{\xt\yt} G^{(i,0)}_{\xt\yt}\right) e^{G_{\xt\yt}}\;,
  \end{split}
\end{align}
\end{subequations}
\intertext{while in the $gg$ channel the various TMDs are given by}
\begin{subequations}\label{eq:finiteNc-tmds-gg}
\begin{align}
\begin{split}
\frac{1}{N_c} \tilde{F}^{(1)}_{gg}(\xt,\yt) =& \f{G^{(i,i)}_{\xt\yt}}{N_c^4 G_{\xt\yt}} \l[\Big((N_c^2-2)N_c^2 G_{\xt\yt} + N_c^2-1\Big) e^{\frac{2N_c^2}{N_c^2-1} G_{\xt\yt}}  -(N_c^2-1)\r]\\
 &\qquad-\frac{N_c^2}{N_c^2-1} G^{(i,0)}_{\xt\yt} G^{(i,0)}_{\xt\yt} e^{\frac{2N_c^2}{N_c^2-1} G_{\xt\yt}}\;,
\end{split}\\
\begin{split}
\frac{1}{N_c} \tilde{F}^{(2)}_{gg}(\xt,\yt) =& \f{G^{(i,i)}_{\xt\yt}}{N_c^4 G_{\xt\yt}} \l[\l(-2N_c^2 G_{\xt\yt} + N_c^2-1\r) e^{\frac{2N_c^2}{N_c^2-1}G_{\xt\yt}}  -(N_c^2-1)\r]\\
&\qquad+ \frac{N_c^2}{N_c^2-1}G^{(i,0)}_{\xt\yt} G^{(i,0)}_{\xt\yt} e^{\frac{2N_c^2}{N_c^2-1}G_{\xt\yt}} \;,
\end{split}\\
\frac{1}{N_c} \tilde{F}^{(3)}_{gg}(\xt,\yt) =& \frac{N_c^2-1}{2N_c^2} G^{(i,i)}_{\xt\yt} \l(e^{\frac{2N_c^2}{N_c^2-1}G_{\xt\yt}}-1\r)\;, \\
\frac{1}{N_c} \tilde{F}^{(4)}_{gg}(\xt,\yt) =& \frac{N_c^2-1}{2N_c^2} G^{(i,i)}_{\xt\yt} \Big(e^{\frac{2N_c^2}{N_c^2-1} G_{\xt\yt}} -1 \Big)\;,\\
\begin{split}
  \frac{1}{N_c} \tilde{F}^{(5)}_{gg}(\xt,\yt)=&\frac{G^{(i,i)}_{\xt\yt}}{8 N_c^2 \left(N_c^2-4\right)G_{\xt\yt}} \Bigg[N_c^3 \left(\left(N_c^3-7 N_c-6\right)e^{\frac{4 N_cG_{\xt\yt}}{N_c+1}}+\left(N_c^3-7 N_c+6\right) e^{\frac{4 N_c G_{\xt\yt}}{N_c-1}}\right)\\
  &\qquad-2\left(N_c^2-4\right)^2 \left(N_c^2-1\right) e^{\frac{4 N_c^2 G_{\xt\yt}}{N_c^2-1}}\Bigg] +\frac{G^{(i,i)}_{\xt\yt}}{8 N_c^2 \left(N_c^2-4\right)G_{\xt\yt}}\\
  &\times\left[-4 \left(N_c^4-13 N_c^2+12\right) e^{\frac{2 N_c^2G_{\xt\yt}}{N_c^2-1}}-4 \left(N_c^2-4\right)\right]
  +\frac{2 N_c^2 G^{(i,0)}_{\xt\yt} G^{(i,0)}_{\xt\yt} e^{\frac{2 N_c^2 G_{\xt\yt}}{N_c^2-1}}}{N_c^2-1}\;,
\end{split}\\
  \begin{split}
      \frac{1}{N_c} \tilde{F}^{(6)}_{gg}(\xt,\yt)=&\frac{G^{(i,i)}_{\xt\yt}}{8 (N_c-2) N_c^4 (N_c+2) G_{\xt\yt}}  \bigg[\left(N_c^3-7 N_c-6\right) N_c^3 e^{\frac{4 N_c G_{\xt\yt}}{N_c+1}}+\left(N_c^3-7 N_c+6\right) N_c^3 e^{\frac{4 N_cG_{\xt\yt}}{N_c-1}}\\
&\qquad+16\left(N_c^2-4\right) N_c^2G_{\xt\yt} e^{\frac{2 N_c^2G_{\xt\yt}}{N_c^2-1}}\bigg]
+\frac{G^{(i,i)}_{\xt\yt}}{8 (N_c-2) N_c^4 (N_c+2)G_{\xt\yt}} \\
&\times\bigg[2 \left(N_c^2-4\right)^2 \left(N_c^2-1\right) e^{\frac{4 N_c^2G_{\xt\yt}}{N_c^2-1}}-4 \left(N_c^6-9 N_c^4+16 N_c^2-8\right) e^{\frac{2 N_c^2G_{\xt\yt}}{N_c^2-1}}\\
&\qquad-4 \left(N_c^2-4\right) N_c^2\bigg] -\frac{2 G^{(i,0)}_{\xt\yt} G^{(i,0)}_{\xt\yt} e^{\frac{2 N_c^2G_{\xt\yt}}{N_c^2-1}}}{N_c^2-1}\;.
\end{split}
\end{align}
\end{subequations}
\end{gather}
Even though in the large $N_c$ limit, the momentum space expressions can be compactly expressed in terms of convolution of various distributions (see e.g.~\cite{MarquetPetreskaRoiesnel16:TMDsFromJIMWLK}), this is no longer the case at finite $N_c$ and the corresponding momentum space expressions have to be obtained by performing the Fourier transformation numerically according to\footnote{Note that for the numerical evaluation of the integral, it is benefitial to split the integration into intervals between zeros of the Bessel function and subsequently sum up all contributions until convergence is reached.}
\begin{eqnarray}
F^{(i)}(\kt) = \frac{4 S_\bot}{g^2 (2\pi)^2} \intd|\xt-\yt|~ |\xt-\yt|~J_{0}(\kt |\xt-\yt|) \tilde{F}^{(i)}(\xt,\yt)\;.
\end{eqnarray}
In Fig.~\ref{fig:TMDs_MV} we present a comparison of numerical lattice results and the Gaussian approximation at large and finite $N_c=3$  for the MV\({}^\gamma\) model, where the Gaussian approximation is exact. Indeed we observe excellent agreement between the different calculations, providing an important cross-check for our calculation. Interestingly, one also observes from this comparison that finite $N_c$ corrections appear to be important for $|\kt|<Q_s$.

% \newpage
\subsection{Calculation of gluon TMDs in Gaussian approximation}
\label{sec:gaussian-tmd-calc}
\subsubsection{Expressions for gluon TMDs in terms of derivatives of local multipole operators}
We relate the position space expressions of the various TMDs to the fundamental Dipole, Quadrupole, Sextupole and Octupole operators
\begin{equation}
S^{(n)}(\xt_{1},\yt_{1},\cdots,\xt_{n},\yt_{n})=\frac{1}{N_c^{n}} \tr \left[  \prod_{i=1}^{n}\V_{\xt_{i}} \Vd_{\yt_{i}}\right]\;,
\end{equation}
such that at the level of a single dipoles $(n=1)$ one obtains the dipole gluon distribution $F^{(1)}_{qg}$ as
\begin{equation}\eq{F1qg-calc}
\frac{1}{N_c} \tilde{F}^{(1)}_{qg}(\xt,\yt)=(+1)~\partial_{i}^{\xt_{1}} \partial_{i}^{\yt_{1}}~\left.S^{(2)}(\xt_{1},\yt_{1})\right|_{\xt_{1}=\xt,\yt_{1}=\yt}\;.
\end{equation}
Similarly, at the level of two dipoles or a quadrupole $(n=2)$ one obtains $F^{(1)}_{gg},F^{(2)}_{gg},F^{(3)}_{gg},F^{(4)}_{gg}$ as
\begin{align}
\frac{1}{N_c} \tilde{F}^{(1)}_{gg}(\xt,\yt)=&(+1)~~~\partial_{i}^{\xt_{1}} \partial_{i}^{\yt_{1}} ~ \left. S^{(2)}(\xt_{1},\yt_{1}) S^{(2)}(\xt_{2},\yt_{2}) \right|_{\substack{\xt_{1}=\xt,\yt_{1}=\yt \\ \xt_{2}=\yt,\yt_{2}=\xt}}\;, \\
\frac{1}{N_c} \tilde{F}^{(2)}_{gg}(\xt,\yt)=&(-1)~~~\partial_{i}^{\xt_{1}} \partial_{i}^{\xt_{2}} ~\left. S^{(2)}(\xt_{1},\yt_{1}) S^{(2)}(\xt_{2},\yt_{2}) \right|_{\substack{\xt_{1}=\xt,\yt_{1}=\yt \\ \xt_{2}=\yt,\yt_{2}=\xt}}\;, \\
\frac{1}{N_c} \tilde{F}^{(3)}_{gg}(\xt,\yt)=\frac{1}{N_c} \tilde{F}^{(4)}_{gg}(\xt,\yt)=&(-N_c)~\partial_{i}^{\xt_{1}} \partial_{i}^{\xt_{2}} \qquad \left. S^{(4)}(\xt_{1},\yt_{2},\xt_{2},\yt_{1}) \right|_{\substack{\xt_{1}=\xt,\yt_{1}=\yt \\ \xt_{2}=\yt,\yt_{2}=\xt}}\;,
\end{align}
while $F^{(2)}_{qg}$ is related to the product of a quadrupole and a dipole $(n=3)$
\begin{equation}
\frac{1}{N_c} \tilde{F}^{(2)}_{qg}(\xt,\yt)=(-N_c)~\partial_{i}^{\xt_{1}} \partial_{i}^{\xt_{2}} \left.S^{(4)}(\xt_{1},\yt_{1},\xt_{2},\yt_{2}) S^{(2)}(\xt_{3},\yt_{3})\right|_{\substack{\xt_{1}=\xt,\yt_{1}=\yt \\ \xt_{2}=\yt,\yt_{2}=\xt \\ \xt_{3}=\yt,\yt_{3}=\xt}}
\end{equation}
and ${F}^{(5)}_{gg}$ are related to the octupol and quadrupole-dipole-dipole operators involving eight Wilson lines $(n=4)$
\begin{align}
\frac{1}{N_c} \tilde{F}^{(5)}_{gg}(\xt,\yt)=& (-N_c^3)~\partial_{i}^{\xt_{1}} \partial_{i}^{\xt_{2}} \left.S^{(8)}(\xt_{1},\yt_{1},\xt_{4},\yt_{4},\xt_{2},\yt_{2},\xt_{3},\yt_{3}) \right|_{\substack{\xt_{1}=\xt,\yt_{1}=\yt \\ \xt_{2}=\yt,\yt_{2}=\xt \\ \xt_{3}=\yt,\yt_{3}=\xt \\ \xt_{4}=\xt,\yt_{4}=\yt }}\;,  \\
\frac{1}{N_c} \tilde{F}^{(6)}_{gg}(\xt,\yt)=& (-N_c)~~\partial_{i}^{\xt_{1}} \partial_{i}^{\xt_{2}}  \left.S^{(4)}(\xt_{1},\yt_{1},\xt_{2},\yt_{2}) S^{(2)}(\xt_{3},\yt_{3}) S^{(2)}(\xt_{4},\yt_{4})\right|_{\substack{\xt_{1}=\xt,\yt_{1}=\yt \\ \xt_{2}=\yt,\yt_{2}=\xt \\ \xt_{3}=\yt,\yt_{3}=\xt \\ \xt_{4}=\xt,\yt_{4}=\yt }}\;,
\end{align}
where in some of the above expressions we used the symmetry $V_{\xt^{i}} \leftrightarrow V^{\dagger}_{\xt^{i}}$ to unify the operator structures.
\subsubsection{Calculating local multipole operators}
% {\Large Basics}
We follow the standard procedure for evaluating the expectation values of the correlation functions of $n$ Wilson lines, by reformulating the problem to a coupled set of evolution equations of various multipole operators. Since the color fields are assumed to follow local Gaussian statistics, the averaging procedure is equivalent to solving the stochastic differential equation
\begin{eqnarray}
\label{eq:WilsonLineEvolutionGaussian}
\V_{\xt}(z+dz)=\V_{\xt}(z)\Big( \id + ig \xi_{\xt}^{a}t^{a}\Big)\;, \qquad  \Vd_{\xt}(z+dz)=\Big( \id - ig \xi_{\xt}^{a}t^{a}\Big) \Vd_{\xt}(z);,
\end{eqnarray}
from $z=0$ to $z=1$ with $V_{\xt}(z=0)=\id$ and Gaussian noise $\xi_{\xt}^{a}(z)$ such that
\begin{eqnarray}
\label{eq:NoiseCorr}
\left< \xi_{\xt}^{a}\xi_{\yt}^{b} \right>=\frac{1}{g^2C_F} \gamma_{\xt\yt}  \delta^{ab} \;.
\end{eqnarray}
Evolution equations for arbitrary products of $2n$ Wilson lines can then be obtained in a straightforward fashion by applying the rules of stochastic calculus. Considering e.g. the simplest possible example of a single dipole ($n=1$) one finds
\begin{equation}
\begin{split}
\bigg\langle \tr \Big[\V_{\xt}(z+dz)&\Vd_{\yt}(z+dz) \Big] \bigg\rangle=\\
 &\left \langle \tr \Big[ \V_{\xt}(z)\Big( \id + ig \xi_{\xt}^{a}t^{a} -\frac{g^2}{2} \xi_{\xt}^{a}t^{a} \xi_{\xt}^{c}t^{c} \Big) \Big( \id - ig \xi_{\yt}^{b}t^{b} - \frac{g^2}{2} \xi_{\yt}^{b}t^{b} \xi_{\yt}^{d}t^{d}\Big) \Vd_{\yt}(z) \Big] \right\rangle
\end{split}
\end{equation}
such that upon evaluating the noise expectation value according to Eq.~(\ref{eq:NoiseCorr}) and using $t^{a}t^{a}=C_F$ to simplify the color structure, one obtains
\begin{eqnarray}
\frac{d}{dz}  \left \langle \tr \Big[ \V_{\xt}(z)\Vd_{\yt}(z)  \Big]\right\rangle =G_{\xt\yt}  \left \langle \tr \Big[ \V_{\xt}(z)\Vd_{\yt}(z)  \Big]\right\rangle\;,
\end{eqnarray}
where we defined the correlation function
\begin{eqnarray}
G_{\xt\yt}=\gamma_{\xt\yt}-\frac{1}{2} \gamma_{\xt\xt} -\frac{1}{2} \gamma_{\yt\yt}\;,
\end{eqnarray}
which frequently appears in color singlet operators. Since we will neglect the impact parameter dependence in the target, the function $G_{\xt\yt}$ is assumed to be a function of only $|\xt-\yt|$, such that
\begin{eqnarray}
G_{\xt\yt}=G(|\xt-\yt|)\;, \qquad G^{(i,0)}_{\xt\yt}= \frac{(\xt-\yt)^{i}}{|\xt-\yt|} G'(|\xt-\yt|)\;, \qquad G^{(0,i)}_{\xt,\yt}= -\frac{(\xt-\yt)^{i}}{|\xt-\yt|} G'(|\xt-\yt|) \\
\end{eqnarray}
while the second derivative is given by
\begin{eqnarray}
 G^{(i,j)}_{\xt\yt}=  -\frac{(\xt-\yt)^{i}}{|\xt-\yt|} \frac{(\xt-\yt)^{j}}{|\xt-\yt|}G''(|\xt-\yt|) - \delta^{ij} \frac{G'(|\xt-\yt|)}{|\xt-\yt|} + \frac{(\xt-\yt)^{i}}{|\xt-\yt|} \frac{(\xt-\yt)^{j}}{|\xt-\yt|}   \frac{G'(|\xt-\yt|)}{|\xt-\yt|}
\end{eqnarray}
such that
\begin{eqnarray}
 G^{(i,i)}_{\xt\yt}=\delta_{ij}  G^{(i,j)}_{\xt\yt} = - \frac{G'(|\xt-\yt|)}{|\xt-\yt|} - G''(|\xt-\yt|)
\end{eqnarray}
which will be relevant for evaluating the TMDs. We also note that by virtue of the definitions the correlation function satisfies the following symmetry properties
\begin{eqnarray}
G_{\xt\xt}=G_{\yt\yt}=0\;, \qquad  G^{(i,0)}_{\xt\xt}= G^{(0,i)}_{\xt\xt}= G^{(i,0)}_{\yt\yt}= G^{(0,i)}_{\yt\yt}=0\,
\end{eqnarray}
\begin{eqnarray}
G_{\xt\yt}=G_{\yt\xt}\;, \qquad G^{(i,0)}_{\yt\xt}= -G^{(i,0)}_{\xt\yt}\;, \qquad G^{(0,i)}_{\yt\xt}=-G^{(0,i)}_{\xt\yt}=G^{(i,0)}_{\xt\yt}\;.
\end{eqnarray}

\subsubsection{Evolution of $n=1$ operators}
We illustrate the procedure at the example of $ \tilde{F}^{(1)}_{qg}(\xt,\yt)$, which according to \refeq{F1qg-calc} is related to the fundamental dipole $S^{(2)}(\xt_{1},\yt_{1})$. By using the prescription in Eq.~(\ref{eq:WilsonLineEvolutionGaussian}) and using the rules of stochastic calculus, it is straightforward to derive the evolution equation for $S^{(2)}(\xt_{1},\yt_{1})$ as
\begin{eqnarray}
\label{eq:EvoEqN1}
\partial_{z} S^{(2)}(\xt_{1},\yt_{1}) = G_{\xt_{1}\yt_{1}} S^{(2)}(\xt_{1},\yt_{1})
\end{eqnarray}
which is trivially solved by  $S^{(2)}(\xt_{1},\yt_{1}|z) =  e^{z G_{\xt_{1}\yt_{1}}}$. By evaluating the expression at $z=1$ and taking the appropriate derivatives, one then immediately obtains
\begin{eqnarray}
\frac{1}{N_c} \tilde{F}^{(1)}_{qg}(\xt,\yt) = \Big(  G^{(i,i)}_{\xt\yt} -  G^{(i,0)}_{\xt\yt} G^{(i,0)}_{\xt\yt}\Big)  e^{G_{\xt\yt}}\;.
\end{eqnarray}
However, from the point of view of evaluating higher order correlation functions $n\geq2$ the exponentiation of the operator on the r.h.s of the evolution equation can be difficult and it is more convenient to evaluate the derivatives prior to exponentiation, by formally expressing the gluon distribution as
\begin{equation}
\begin{split}
\label{eq:ExpDerivative}
\frac{1}{N_c} \tilde{F}^{(1)}_{qg}(\xt,\yt) =& \int_{0}^{1} ds~e^{s G_{\xt_{1}\yt_{1}} }~G^{(i,i)}_{\xt_{1}\yt_{1}}~e^{(1-s) G_{\xt_{1}\yt_{1}} } \\
&+ \int_{0}^{1} ds~s\int_{0}^{1} dt ~e^{s t G_{\xt_{1}\yt_{1}} }~G^{(i,0)}_{\xt_{1}\yt_{1}}~e^{s (1-t) G_{\xt_{1}\yt_{1}} }~G^{(0,i)}_{\xt_{1}\yt_{1}}~e^{(1-s) G_{\xt_{1}\yt_{1}} } \\
&+ \left. \int_{0}^{1} ds~(1-s)\int_{0}^{1} dt ~e^{sG_{\xt_{1}\yt_{1}} }~G^{(0,i)}_{\xt_{1}\yt_{1}}~e^{(1-s) t G_{\xt_{1}\yt_{1}} }~G^{(i,0)}_{\xt_{1}\yt_{1}}~e^{(1-s)(1-t) G_{\xt_{1}\yt_{1}} } \right|_{\substack{\xt_{1}=\xt,\\ \yt_{1}=\yt}}\;,
\end{split}
\end{equation}
which obviously gives the same results upon integration.

\subsubsection{Evolution of $n=2$ operators}
Clearly, this strategy becomes advantageous already in the case of $n=2$ relevant for the calculation of $F^{(1)}_{gg},F^{(2)}_{gg},F^{(3)}_{gg},F^{(4)}_{gg}$. By following the procedure outlined above, one obtains a coupled set of evolution equations for the dipole-dipole and quadrupole operators
\begin{equation}
\begin{split}
\label{eq:EvoEqN2}
\partial_{z}\!\begin{pmatrix}\! S^{(2)}(\xt_{1},\yt_{1})  S^{(2)}(\xt_{2},\yt_{2}) \\  S^{(4)}(\xt_{1},\yt_{2},\xt_{2},\yt_{1})\!\end{pmatrix}\!=&\!\begin{pmatrix}\! G_{\xt_{1}\yt_{1}}\!\!+\!G_{\xt_{2}\yt_{2}}\!\!-\frac{T_{\xt_{1}\yt_{1} \leftrightarrow \xt_{2}\yt_{2}}}{N_c^2-1}  & \frac{1}{2C_F} T_{\xt_{1}\yt_{1} \leftrightarrow \xt_{2}\yt_{2}}  \\  T_{\xt_{1}\yt_{2} \leftrightarrow \xt_{2}\yt_{1}} &  G_{\xt_{1}\yt_{2}}\!\!+\!G_{\xt_{2}\yt_{1}}\!\!- \frac{T_{\xt_{1}\yt_{2} \leftrightarrow \xt_{2}\yt_{1}}}{N_c^2-1} \! \end{pmatrix}\\
  &\times\begin{pmatrix} S^{(2)}(\xt_{1},\yt_{1})  S^{(2)}(\xt_{2},\yt_{2}) \\  S^{(4)}(\xt_{1},\yt_{2},\xt_{2},\yt_{1}) \end{pmatrix}
\end{split}
\end{equation}
where the transition matrix element takes the form
\begin{eqnarray}
T_{\xt_{1}\yt_{1} \leftrightarrow \xt_{2}\yt_{2}}  = G_{\xt_{1}\yt_{2}} + G_{\xt_{2}\yt_{1}} - G_{\xt_{1}\xt_{2}} - G_{\yt_{1}\yt_{2}}\;.
\end{eqnarray}
Even though it is still possible in this case to perform the exponentiation of the operator on the r.h.s. (see e.g. [XXX]) it is much easier to first take the derivatives and set the coordinates to the desired values
% \begin{eqnarray}
\begin{equation}
\begin{split}
\begin{pmatrix} G_{\xt_{1}\yt_{1}} + G_{\xt_{2}\yt_{2}} -\frac{T_{\xt_{1}\yt_{1} \leftrightarrow \xt_{2}\yt_{2}}}{N_c^2-1} & \frac{1}{2C_F} T_{\xt_{1}\yt_{1} \leftrightarrow \xt_{2}\yt_{2}}  \\  T_{\xt_{1}\yt_{2} \leftrightarrow \xt_{2}\yt_{1}} &  G_{\xt_{1}\yt_{2}} + G_{\xt_{2}\yt_{1}} - \frac{T_{\xt_{1}\yt_{2} \leftrightarrow \xt_{2}\yt_{1}}}{N_c^2-1}    \end{pmatrix}_{{\Bigg|}_{\substack{\xt_{1}=\xt,\yt_{1}=\yt \\ \xt_{2}=\yt,\yt_{2}=\xt}}}\!\!\!\!\!\!\!\!\!\!=\!\!
\begin{pmatrix} \frac{2N_c^2}{N_c^2-1} G_{\xt\yt} & -\frac{1}{C_F}  G_{\xt\yt} \\ 0 & 0 \end{pmatrix}
\end{split}
\end{equation}
% \end{eqnarray}
which is easily exponentiated. By taking the various derivatives of the evolution operator on the r.h.s of Eq.~(\ref{eq:EvoEqN2}), the gluon TMDs $F^{(1)}_{gg},F^{(2)}_{gg}$ can then be evaluated as in Eq.~(\ref{eq:ExpDerivative}) yielding
\begin{align}
\begin{split}
\frac{1}{N_c} \tilde{F}^{(1)}_{gg}(\xt,\yt) =& \frac{G^{(i,i)}_{\xt\yt}}{N_c^4 G_{\xt\yt}} \l[\Big((N_c^2-2)N_c^2 G_{\xt\yt} + N_c^2-1\Big) e^{\frac{2N_c^2}{N_c^2-1} G_{\xt\yt}}  -(N_c^2-1)\r] \\
&\qquad-\frac{N_c^2}{N_c^2-1} G^{(i,0)}_{\xt\yt} G^{(i,0)}_{\xt\yt} e^{\frac{2N_c^2}{N_c^2-1} G_{\xt\yt}}\;,
\end{split} \\
\begin{split}
\frac{1}{N_c} \tilde{F}^{(2)}_{gg}(\xt,\yt) =&\f{G^{(i,i)}_{\xt\yt}}{N_c^4 G_{\xt\yt}} \l[\Big(-2N_c^2 G_{\xt\yt} + N_c^2-1\Big) e^{\frac{2N_c^2}{N_c^2-1} G_{\xt\yt}}  -(N_c^2-1)\r] \\
&\qquad+\frac{N_c^2}{N_c^2-1} G^{(i,0)}_{\xt\yt} G^{(i,0)}_{\xt\yt} e^{\frac{2N_c^2}{N_c^2-1} G_{\xt\yt}}\;.
\end{split}
\end{align}
Similarly one also finds the well known result for the Weiszaecker-Williams distribution $F^{(3)}_{gg},F^{(4)}_{gg}$ [{color{red}XXX}], which in our notation takes the form
\begin{eqnarray}
\frac{1}{N_c} \tilde{F}^{(3)}_{gg}(\xt,\yt) &=& \frac{1}{N_c} \tilde{F}^{(4)}_{gg}(\xt,\yt) = \frac{N_c^2-1}{2N_c^2} G^{(i,i)}_{\xt\yt} \Big(e^{\frac{2N_c^2}{N_c^2-1} G_{\xt\yt}} -1 \Big)\;.
\end{eqnarray}

\subsubsection{Evolution of general operators}
By using the prescription in Eq.~(\ref{eq:WilsonLineEvolutionGaussian}) it is straightforward to obtain the evolution equations for arbitrary products of $2n$ Wilson lines. Considering the most general operator consisting of $2n$ Wilson lines
\begin{eqnarray}
O(\xt_{1},\yt_{1},\cdots,\xt_{n},\yt_{n}) =\frac{1}{N_c^{n}} \prod_{m=1}^{M} \tr \left[ \prod_{j=1}^{n_{m}}\V_{\xt_{m_{j}}} \Vd_{\yt_{m_{j}}}  \right]\;,
\end{eqnarray}
one has a number of different possibilities for the insertion of the Gaussian noise $\xi^{a} t^{a}$, which upon performing the Gaussian averaging schematically correspond to a) the one possible insertion of generators between neighboring Wilson lines at position $q$ of operator number $p$
\begin{align}
\begin{split}
\partial_{z}& O(\xt_{1},\yt_{1},\cdots,\xt_{n},\yt_{n}) \big|_{a}\!\\
&=\!\frac{1}{N_c^{n}}   \sum_{p=1}^{M} \sum_{q=1}^{n_{p}} \frac{G_{\xt_{p_{q}}\yt_{p_{q}}}}{C_F}
\tr\!\left[\left(\prod_{j=1}^{q-1}\!\V_{\xt_{p_{j}}} \!\Vd_{\yt_{p_{j}}} \right) \!\V_{\xt_{p_{q}}} t^{a} t^{a} \Vd_{\yt_{p_{q}}} \left(\prod_{j=q+1}^{n_{p}}\!\!\V_{\xt_{p_{j}}} \!\Vd_{\yt_{p_{j}}}\!\right)\!\right]\prod_{\substack{m=1,\\m\neq p}}^{M}\!\!\tr\!\left[ \prod_{j=1}^{n_{m}}\!\V_{\xt_{m_{j}}}\!\Vd_{\yt_{m_{j}}}\!\right]
\end{split}\n \\
&\qquad\qquad\qquad\qquad= \left(\sum_{p=1}^{M} \sum_{q=1}^{n_{m}}  G_{\xt_{p_{q}}\yt_{p_{q}}} \right) O(\xt_{1},\yt_{1},\cdots,\xt_{n},\yt_{n})\;,
\end{align}
which leaves the operator structure intact. Or b) the four possible insertions of generators between non-neighboring Wilson lines at position $q$ and $l>q$ of the same operator number $p$
\begin{equation}
\begin{split}
&\partial_{z} O(\xt_{1},\yt_{1},\cdots,\xt_{n},\yt_{n}) \big|_{b}
\!\!=\!\frac{1}{N_c^{n}} \sum_{p=1}^{M}\!\sum_{q=1}^{n_{p}-1}\!\sum_{l=q+1}^{n_{p}}  \frac{G_{\xt_{p_{q}}\yt_{p_{l}}}\!\!+\!G_{\xt_{p_{l}}\yt_{p_{q}}}\!\!-\! G_{\xt_{p_{q}}\xt_{p_{l}}}\!\!-\!G_{\yt_{p_{q}}\yt_{p_{l}}}}{C_F} \\
&\quad\times\!\tr\!\!\left[\!\left(\prod_{j=1}^{q-1}\!\V_{\xt_{p_{j}}} \!\Vd_{\yt_{p_{j}}} \!\right) \!\V_{\xt_{p_{q}}} t^{a} \Vd_{\yt_{p_{q}}} \!\left(\prod_{j=q+1}^{l-1}\!\!\V_{\xt_{p_{j}}} \!\Vd_{\yt_{p_{j}}} \!\right) \!\V_{\xt_{p_{l}}} t^{a} \Vd_{\yt_{p_{l}}}  \!\left(\prod_{j=l+1}^{n_{p}}\!\!\V_{\xt_{p_{j}}} \!\Vd_{\yt_{p_{j}}} \!\right)  \!\right]\! \prod_{\substack{m=1,\\m\neq p}}^{M}\!\!\!\tr\!\left[ \prod_{j=1}^{n_{m}}\!\V_{\xt_{m_{j}}} \!\Vd_{\yt_{m_{j}}} \! \right]
\end{split}
\end{equation}
By using the Fierz identity $t^{a}_{ij} t^{a}_{kl}= \frac{1}{2} \delta_{il} \delta_{kj} - \frac{1}{2N_c} \delta_{ij} \delta_{kl}$, the expression can be further simplified and separated into two distinct contributions. While the first contribution generates an additional trace, the second term proportional to the identity does not change the original operator structure. By adding the two contributions one obtains
\begin{equation}
\begin{split}
  \partial_{z}O(\xt_{1},\yt_{1},\cdots,\xt_{n},\yt_{n})\big|_{b}\!=&\!\frac{1}{N_c^{n}}   \sum_{p=1}^{M} \sum_{q=1}^{n_{p}-1}  \sum_{l=q+1}^{n_{p}} \! T_{\xt_{p_{q}}\yt_{p_{q}} \leftrightarrow \xt_{p_{l}}\yt_{p_{l}}} \\
  &\quad\times\!\tr\!\left[\!\left(\prod_{j=1}^{q-1}\!\V_{\xt_{p_{j}}} \!\Vd_{\yt_{p_{j}}}\! \right) \!\V_{\xt_{p_{q}}}  \!\Vd_{\yt_{p_{l}}}  \!\left(\prod_{j=l+1}^{n_{p}}\!\!\V_{\xt_{p_{j}}} \!\Vd_{\yt_{p_{j}}} \right)\!\right]\\
  &\quad\times\!\tr\!\left[\!\Vd_{\yt_{p_{q}}} \!\left(\prod_{j=q+1}^{l-1}\!\!\V_{\xt_{p_{j}}} \!\Vd_{\yt_{p_{j}}} \right) \!\V_{\xt_{p_{l}}}  \right]
\!\!\prod_{\substack{m=1,\\m\neq p}}^{M}\!\tr\!\left[\!\prod_{j=1}^{n_{m}}\!\!\V_{\xt_{m_{j}}} \!\Vd_{\yt_{m_{j}}}  \right] \\
& -\frac{1}{N_c^2-1} \!\left(\sum_{p=1}^{M} \sum_{q=1}^{n_{p}-1}  \sum_{l=q+1}^{n_{p}}  T_{\xt_{p_{q}}\yt_{p_{q}} \leftrightarrow \xt_{p_{l}}\yt_{p_{l}}}\right)  O(\xt_{1},\yt_{1},\cdots,\xt_{n},\yt_{n})
\end{split}
\end{equation}
where we defined the transition matrix element
\begin{eqnarray}
T_{\xt_{p_{q}}\yt_{p_{q}} \leftrightarrow \xt_{p_{l}}\yt_{p_{l}}} = G_{\xt_{p_{q}}\yt_{p_{l}}} + G_{\xt_{p_{l}}\yt_{p_{q}}} - G_{\xt_{p_{q}}\xt_{p_{l}}} - G_{\yt_{p_{q}}\yt_{p_{l}}}\;.
\end{eqnarray}
Of course, there are also c)  the four possible insertions of generators between Wilson lines at position $q$ and $l$ of the different operators numbered $p$ and $k>p$
\begin{equation}
\begin{split}
\partial_{z}& O(\xt_{1},\yt_{1},\cdots,\xt_{n},\yt_{n}) \big|_{c}
\!=\!\frac{1}{N_c^{n}}   \sum_{p=1}^{M} \sum_{k=p+1}^{M} \sum_{q=1}^{n_{p}}  \sum_{l=1}^{n_{k}}  \frac{T_{\xt_{p_{q}}\yt_{p_{q}} \leftrightarrow \xt_{k_{l}}\yt_{k_{l}}}}{C_F}  \!\left(\!\prod_{\substack{m=1,\\ m\neq p,k}}^{M}\!\!\tr\!\left[ \prod_{j=1}^{n_{m}}\!\V_{\xt_{m_{j}}}\!\Vd_{\yt_{m_{j}}} \!\right]\!\right) \\
&\times\!\tr\!\left[\!\left(\prod_{j=1}^{q-1}\!\V_{\xt_{p_{j}}}\!\Vd_{\yt_{p_{j}}}\! \right) \!\V_{\xt_{p_{q}}} t^{a} \Vd_{\yt_{p_{q}}}\! \left(\prod_{j=q+1}^{n_{p}}\!\!\V_{\xt_{p_{j}}} \!\Vd_{\yt_{p_{j}}}\!\right) \!\right]
\!\!\tr\!\left[\!\left(\prod_{j=1}^{l-1}\!\V_{\xt_{k_{j}}}\!\Vd_{\yt_{k_{j}}}\!\right) \!\V_{\xt_{k_{l}}} t^{a} \Vd_{\yt_{k_{l}}}\!\left(\prod_{j=l+1}^{n_{k}}\!\!\V_{\xt_{k_{j}}} \!\Vd_{\yt_{k_{j}}}\! \right)  \right]
\end{split}
\end{equation}
which following the same steps as indicated above yields
\begin{equation}
\begin{split}
&\left.\partial_{z} O(\xt_{1},\yt_{1},\cdots,\xt_{n},\yt_{n}) \right|_{c}  \\
&= \frac{1}{N_c^{n}}   \sum_{p=1}^{M} \sum_{k=p+1}^{M} \sum_{q=1}^{n_{p}}  \sum_{l=1}^{n_{k}}  \frac{T_{\xt_{p_{q}}\yt_{p_{q}} \leftrightarrow \xt_{k_{l}}\yt_{k_{l}}}}{C_F}  \left(\prod_{\substack{m=1,\\ m\neq p,k}}^{M} \tr \left[ \prod_{j=1}^{n_{m}}\V_{\xt_{m_{j}}} \Vd_{\yt_{m_{j}}}  \right]\right)  \\
&\quad\times\! \tr\! \left[\! \left(\prod_{j=1}^{q-1}\!\V_{\xt_{p_{j}}} \!\Vd_{\yt_{p_{j}}} \right) \!\V_{\xt_{p_{q}}}\!\Vd_{\yt_{k_{l}}} \left(\prod_{j=l+1}^{n_{k}}\!\V_{\xt_{k_{j}}} \!\Vd_{\yt_{k_{j}}} \right) \left(\prod_{j=1}^{l-1}\!\V_{\xt_{k_{j}}} \!\Vd_{\yt_{k_{j}}} \right) \!\V_{\xt_{k_{l}}} \!\Vd_{\yt_{p_{q}}} \left(\prod_{j=q+1}^{n_{p}}\!\V_{\xt_{p_{j}}} \!\Vd_{\yt_{p_{j}}} \right)  \right]\\
&\quad-\frac{1}{N_c^2-1}  \left(\sum_{p=1}^{M} \sum_{k=p+1}^{M} \sum_{q=1}^{n_{p}}  \sum_{l=1}^{n_{k}} T_{\xt_{p_{q}}\yt_{p_{q}} \leftrightarrow \xt_{k_{l}}\yt_{k_{l}}}\right) O(\xt_{1},\yt_{1},\cdots,\xt_{n},\yt_{n})
\end{split}
\end{equation}
Even though the general expressions are rather lengthy, and the number of distinct operators grows rapidly as a function of $n$ the procedure can be automated in a straightforward way and is easily tractable for the $n$-values of interest. One first constructs a basis of all possible configurations. Starting from a configuration of $n$ fundamental dipoles $\prod_{i=1}^{n} S^{(2)}(\xt_{i},\yt_{i})$, one considers all possible generator insertions a),b) and c) and add the new operator structures to the basis. By repeating the second step until no new operators are generated ($n-1$ times) one obtains a complete basis for the evolution of the $2n$ Wilson line correlators. Next one constructs the elements of the evolution matrix, by considering all possible generator insertions a),b) and c) on each basis state. By identifying the relevant operators for the TMD calculation, one can then take the relevant derivatives and subsequently set the coordinates $(\xt_{1},\yt_{1},\cdots,\xt_{n},\yt_{n})$ to $\xt,\yt$ as required for the evaluation of the operator. By performing the exponentiation as in Eq.~(\ref{eq:ExpDerivative}) and projecting the result on the appropriate operator configuration one obtains the expression for the TMD. Since the expressions involved can be rather lengthy, we refrain from providing further details on the intermediate steps of the calculation and simply quote the final results.
\begin{align}
\begin{split}
\frac{1}{N_c} \tilde{F}^{(2)}_{gq}(\xt,\yt)&= \left( (N_c+2)(N_c-1)e^{\frac{3N_c-1}{N_c^2-1} G_{\xt\yt}} + (N_c-2)(N_c+1)^2 e^{\frac{3N_c+1}{N_c^2-1} G_{\xt\yt}} -2N_c(N_c^2-3)e^{G_{\xt\yt}}\right)\\
&\times\frac{G^{(i,i)}_{\xt\yt}}{4N_c(N_c^2-1)G_{\xt\yt}}
+ \frac{1}{N_c^2-1} \left( G^{(i,i)}_{\xt\yt} - G^{(i,0)}_{\xt\yt} G^{(i,0)}_{\xt\yt}\right) e^{G_{\xt\yt}}
\end{split}\\
\begin{split}
\frac{1}{N_c} \tilde{F}^{(5)}_{gg}(\xt,\yt)&=\frac{G^{(i,i)}_{\xt\yt} }{8 N_c^2 \left(N_c^2-4\right)G_{\xt\yt}}\Bigg[N_c^3 \left(\left(N_c^3-7 N_c-6\right)e^{\frac{4 N_cG_{\xt\yt}}{N_c+1}}+\left(N_c^3-7 N_c+6\right) e^{\frac{4 N_c
  G_{\xt\yt}}{N_c-1}}\right)\\
&\qquad-2 \left(N_c^2-4\right)^2 \left(N_c^2-1\right) e^{\frac{4 N_c^2
    G_{\xt\yt}}{N_c^2-1}}\Bigg]+\frac{G^{(i,i)}_{\xt\yt}}{8 N_c^2 \left(N_c^2-4\right)G_{\xt\yt}} \\
&\quad\qquad\times\left(-4 \left(N_c^4-13 N_c^2+12\right) e^{\frac{2 N_c^2G_{\xt\yt}}{N_c^2-1}}-4
   \left(N_c^2-4\right)\right) + \frac{2 N_c^2 G^{(i,0)}_{\xt\yt} G^{(i,0)}_{\xt\yt} e^{\frac{2 N_c^2 G_{\xt\yt}}{N_c^2-1}}}{N_c^2-1}
\end{split}\\
%
% \end{eqnarray}
\begin{split}
\frac{1}{N_c} \tilde{F}^{(6)}_{gg}(\xt,\yt)&=
\frac{G^{(i,i)}_{\xt\yt}}{8 (N_c-2) N_c^4 (N_c+2) G_{\xt\yt}} \Bigg[\!\left(N_c^3-7 N_c-6\right) N_c^3 e^{\frac{4 N_c G_{\xt\yt}}{N_c+1}}\!\!+\!\left(N_c^3-7 N_c+6\right) N_c^3 e^{\frac{4 N_cG_{\xt\yt}}{N_c-1}}\\
&\qquad+16 \left(N_c^2-4\right) N_c^2G_{\xt\yt} e^{\frac{2 N_c^2G_{\xt\yt}}{N_c^2-1}}\Bigg]+\frac{G^{(i,i)}_{\xt\yt} }{8 (N_c-2) N_c^4 (N_c+2)G_{\xt\yt}} \\
  &\quad\quad\times\Bigg[2 \left(N_c^2-4\right)^2 \left(N_c^2-1\right) e^{\frac{4 N_c^2G_{\xt\yt}}{N_c^2-1}}-4 \left(N_c^6-9 N_c^4+16 N_c^2-8\right) e^{\frac{2 N_c^2G_{\xt\yt}}{N_c^2-1}}\\
&\qquad\qquad-4 \left(N_c^2-4\right) N_c^2\Bigg] -\frac{2 G^{(i,0)}_{\xt\yt} G^{(i,0)}_{\xt\yt} e^{\frac{2 N_c^2G_{\xt\yt}}{N_c^2-1}}}{N_c^2-1}
\end{split}
\end{align}

% \newpage
%%%%%%%%% LATTICE %%%%%%%%%%%%%%
\subsection{Calculation of Gluon TMD's on the lattice}
\label{sec:lattice-tmd-calc}
Below we explain the calculation of the various gluon TMDs from the Wilson line configurations $\V_{\xt}$ discretized on a 2-D transverse lattice, as it is standard for studying e.g. JIMWLK evolution. We discretize the light-cone fields $E_{\mu,\xt}^{(+/-)}$
\begin{eqnarray}
E_{\mu,\xt}^{(+)}=\V_{\xt} \Big(i\partial_{\mu}\Vd_{\xt}\Big)\; \qquad E_{\mu,\xt}^{(-)}= \Big(i\partial_{\mu}\Vd_{\xt}\Big) \V_{\xt}\; \qquad E_{\mu,\xt}^{(-)}=V^{\dagger}_{\xt} E_{\mu,\xt}^{(+)} V_{\xt}\;.
\end{eqnarray}
according to the standard lattice plaquette definition
\begin{subequations}
\begin{align}
E^{(+)}_{\mu,\xt}=&E^{(+)a}_{\mu,\xt}~t^{a}\;, & E^{(+),a}_{\mu,\xt}=\ReTr\Big[it^{a}\Big(\V_{\xt}\Vd_{\xt+\mu} +\V_{\xt-\mu}\Vd_{\xt} \Big)\Big]\;,  \\
E^{(-)}_{\mu,\xt}=&E^{(-)a}_{\mu,\xt}~t^{a}\;, & E^{(-),a}_{\mu,\xt}=\ReTr\Big[it^{a}\Big(\Vd_{\xt+\mu}\V_{\xt} +\Vd_{\xt}\V_{\xt-\mu} \Big)\Big]\;,
\end{align}
\end{subequations}
which explicitly ensures the tracelessness and hermiticity properties of the $SU(N_c)$ algebra valued fields $E_{\mu,\xt}^{(+/-)}$. By virtue of the $SU(N_c)$ Fierz identity $t^{a}_{ij} t^{a}_{kl}= \frac{1}{2} \delta_{il} \delta_{kj} - \frac{1}{2N_c} \delta_{ij} \delta_{kl}$ one then has
\begin{subequations}
\begin{align}
\begin{split}
  E^{(+)}_{\mu,\xt}&=\frac{i}{4} \left[ \V_{\xt}\Vd_{\xt+\mu} + \V_{\xt-\mu}\Vd_{\xt} - \V_{\xt+\mu} \Vd_{\xt} - \V_{\xt}\Vd_{\xt-\mu}\right]\\
  &\qquad-\frac{i}{4N_c} \tr \left[ \V_{\xt}\Vd_{\xt+\mu} + \V_{\xt-\mu}\Vd_{\xt} - \V_{\xt+\mu} \Vd_{\xt} - \V_{\xt}\Vd_{\xt-\mu}\right]\;,
\end{split}\\
\begin{split}
E^{(-)}_{\mu,\xt}&=\frac{i}{4} \left[ \Vd_{\xt+\mu}\V_{\xt} +\Vd_{\xt}\V_{\xt-\mu} - \Vd_{\xt} \V_{\xt+\mu} -\Vd_{\xt-\mu} \V_{\xt}\right]\\
  &\qquad-\frac{i}{4N_c} \tr \left[ \Vd_{\xt+\mu}\V_{\xt} +\Vd_{\xt}\V_{\xt-\mu} - \Vd_{\xt} \V_{\xt+\mu} -\Vd_{\xt-\mu} \V_{\xt}\right]\;.
\end{split}
\end{align}
\end{subequations}
We note that based on the unitarity relation
\begin{equation}
\Big(\partial_{\mu}\V_{\xt}\Big)\V^{\dagger}_{\xt} = -\V_{\xt} \Big(\partial_{\mu}\Vd_{\xt}\Big)
\end{equation}
 we can then express all derivative of Wilson lines according to
\begin{equation}
\Big(\partial_{\mu}\Vd_{\xt}\Big)=(-i)\Vd_{\xt} E_{\mu,\xt}^{(+)}= (-i) E_{\mu,\xt}^{(-)} \V_{\xt}\;, \qquad
\Big(\partial_{\mu}\V_{\xt}\Big)=(+i) E_{\mu,\xt}^{(+)}~\V_{\xt} = (+i) \V_{\xt} E_{\mu,\xt}^{(-)} \;,
\end{equation}
such that all relevant correlation functions can be expressed solely in terms of products of light-like Wilson lines $\V_{\xt}$ and light-cone fields $E_{\mu,\xt}^{(+/-)}$. Based on the above defintions of lattice operators, the small-x TMDs in the $qg$ channel  can be compactly expressed as
% \red{Perhaps we can omit the $e^{-ik(x-y)}$ and simply give the expressions for $\F$ instead of $F$}
\begin{gather}
\begin{subequations}
\begin{align}
\begin{split}
\frac{g^2 (2\pi)^3}{4 a_s^2} F^{(1)}_{qg}(\kt)
&=\sum_{\xt,\yt} e^{-i\kt(\xt-\yt)}~\tr\left[ \Vd_{\xt}~E^{(+)}_{\mu,\xt}~E^{(+)}_{\mu,\yt}~\V_{\yt} \right]\\
&= \sum_{\xt,\yt} e^{-i\kt(\xt-\yt)}~\tr\left[E^{(-)}_{\mu,\xt}~\Vd_{\xt}~\V_{\yt}~E^{(-)}_{\mu,\yt} \right]\;,
\end{split} \\
  \begin{split}
\frac{g^2 (2\pi)^3}{4 a_s^2}  F^{(2)}_{qg}(\kt)
&=\frac{1}{N_c}~\sum_{\xt,\yt} e^{-i\kt(\xt-\yt)}~\tr\left[ \Vd_{\xt} E^{(+)}_{\mu,\xt} V_{\xt} \Vd_{\yt}E^{(+)}_{\mu,\yt} V_{\yt}\right] \tr\Big[\Vd_{\xt}\V_{\yt} \Big]\\
&=\frac{1}{N_c}~\sum_{\xt,\yt} e^{-i\kt(\xt-\yt)}~\tr\left[ E^{(-)}_{\mu,\xt}~E^{(-)}_{\mu,\yt}\right] \tr\Big[\Vd_{\xt}\V_{\yt} \Big]\;,
\end{split}
\end{align}
\end{subequations}
\intertext{and similary one finds the following expressions for the small-x TMD's in the $gg$ channel}
\begin{subequations}
\begin{align}
  \begin{split}
\frac{g^2 (2\pi)^3}{4 a_s^2}  F^{(1)}_{gg}(\kt)
&=\frac{1}{N_c}~\sum_{\xt,\yt} e^{-i\kt(\xt-\yt)}~\tr\left[ \Vd_{\xt}~E_{\mu,\xt}^{(+)}~E^{(+)}_{\mu,\yt}~\V_{\yt} \right]  \tr\Big[\V_{\xt}\Vd_{\yt} \Big]\\
&=\frac{1}{N_c}~\sum_{\xt,\yt} e^{-i\kt(\xt-\yt)}~\tr\left[ E_{\mu,\xt}^{(-)}\Vd_{\xt}~\V_{\yt}E^{(-)}_{\mu,\yt} \right]  \tr\Big[\V_{\xt}\Vd_{\yt} \Big]
\end{split}\\
  \begin{split}
\frac{g^2 (2\pi)^3}{4 a_s^2}  F^{(2)}_{gg}(\kt)
&=\frac{1}{N_c}~\sum_{\xt,\yt} e^{-i\kt(\xt-\yt)}~\tr\left[ E_{\mu,\xt}^{(+)}~\V_{\xt}~\Vd_{\yt}\right] \tr\left[E_{\mu,\yt}^{(+)}~\V_{\yt}~\Vd_{\xt}\right]\\
&=\frac{1}{N_c}~\sum_{\xt,\yt} e^{-i\kt(\xt-\yt)}~\tr\left[ \V_{\xt}E_{\mu,\xt}^{(-)}~\Vd_{\yt}\right] \tr\left[\V_{\yt}E_{\mu,\yt}^{(-)}~\Vd_{\xt}\right]
\end{split}\\
  \begin{split}
\frac{g^2 (2\pi)^3}{4 a_s^2}  F^{(3/4)}_{gg}(\kt)
&= \sum_{\xt,\yt} e^{-i\kt(\xt-\yt)}~\tr\left[ E_{\mu,\xt}^{(+)} E_{\mu,\yt}^{(+)}\right]\\
&= \sum_{\xt,\yt} e^{-i\kt(\xt-\yt)}~\tr\left[ \V_{\xt}E_{\mu,\xt}^{(-)} \Vd_{\xt}~\V_{\yt}E_{\mu,\yt}^{(-)} \Vd_{\yt}\right]
\end{split}\\
  \begin{split}
\frac{g^2 (2\pi)^3}{4 a_s^2}  F^{(5)}_{gg}(\kt)
&=\sum_{\xt,\yt} e^{-i\kt(\xt-\yt)}~\tr\left[ \Vd_{\xt} E_{\mu,\xt}^{(+)} \V_{\xt}\Vd_{\yt}~\V_{\xt}\Vd_{\yt} E_{\mu,\yt}^{(+)} V_{\yt}\Vd_{\xt}\V_{\yt}\right] \\
&= \sum_{\xt,\yt} e^{-i\kt(\xt-\yt)}\tr\left[ E_{\mu,\xt}^{(-)}\Vd_{\yt}\V_{\xt} E_{\mu,\yt}^{(-)}\Vd_{\xt}\V_{\yt}\right]
\end{split}\\
    \begin{split}
\frac{g^2 (2\pi)^3}{4 a_s^2}  F^{(6)}_{gg}(\kt)
&=\frac{1}{N_c^2}\sum_{\xt,\yt} e^{-i\kt(\xt-\yt)}~\tr\left[ \Vd_{\xt} E_{\mu,\xt}^{(+)} \V_{\xt}\Vd_{\yt} E_{\mu,\yt}^{(+)} V_{\yt}\right]~\tr[\V_{\xt} \Vd_{\yt}]\tr[\V_{\yt}\Vd_{\xt}]\\\
&=\frac{1}{N_c^2}~\sum_{\xt,\yt} e^{-i\kt(\xt-\yt)}~\tr\left[E_{\mu,\xt}^{(-)}E_{\mu,\yt}^{(-)} V_{\yt}\right]\tr[\V_{\xt} \Vd_{\yt}]\tr[\V_{\yt}\Vd_{\xt}]
\end{split}
\end{align}
\end{subequations}
\end{gather}
Notably, the definitions in terms of $E_{\mu,\xt}^{(-)}$ are simpler to implement, as they require only a single operator structure for each channel. Specifically, by defining operators with open color indices according to \red{XXX} for the $qg$ channel
\begin{subequations}
\begin{align}
\left( E_{\mu}^{(-)} \Vd  \right)^{ij,kl}_{\kt} &= \sum_{\xt} e^{-i\kt \xt} \Big(E_{\mu,\xt}^{(-)}\Big)^{ij} \Big(\Vd_{\xt}\Big)^{kl}\;,  \\
\left( E_{\mu}^{(-)} \V  \right)^{ij,kl}_{\kt} &= \sum_{\xt} e^{-i\kt \xt} \Big(E_{\mu,\xt}^{(-)}\Big)^{ij} \Big(\V_{\xt}\Big)^{kl}\;,
\end{align}
\end{subequations}
one can make use of the identities $\V^{kl}=\Big(\big(V^{\dagger}\big)^{lk}\Big)^{*}$ and $\Big(E_{\mu}^{(-)}\Big)^{ij} = \Big(\Big(E_{\mu,\xt}^{(-)}\Big)^{ji}\Big)^{*}$ to express
\begin{equation}
\left( E_{\mu}^{(-)} \V  \right)^{ij,kl}_{-\kt}=  \Big(\sum_{\xt} e^{-i\kt \xt} \Big(E_{\mu,\xt}^{(-)}\Big)^{ji} \Big(\Vd_{\xt}\Big)^{lk} \Big)^{*} = \left( E_{\mu}^{(-)} \Vd  \right)^{ji,lk}_{\kt}\;.
\end{equation}
Based on this procedure, the small-x gluon TMDs in the $qg$ channel are then given by
\begin{subequations}
\begin{align}
\frac{g^2 (2\pi)^3}{4 a_s^2} F^{(1)}_{qg}(\kt) &=  \left| \left( E_{\mu}^{(-)} \Vd  \right)^{ik,kj}_{\kt}  \right|^2 \;, \\
\frac{g^2 (2\pi)^3}{4 a_s^2} F^{(2)}_{qg}(\kt) &=  \frac{1}{N_c}~\left| \left( E_{\mu}^{(-)} \Vd  \right)^{ij,kl}_{\kt}  \right|^2\;.
\end{align}
\end{subequations}
Similarly, we can define a single operator with open color indices for the $gg$ channel according to
\begin{equation}
\left( E_{\mu}^{(-)} \Vd \V \right)^{ij,kl,mn}_{\kt} = \sum_{\xt} e^{-i\kt \xt} \Big(E_{\mu,\xt}^{(-)}\Big)^{ij} \Big(\Vd_{\xt}\Big)^{kl} \Big( \V_{\xt} \Big)^{mn}\;,
\end{equation}
which has the following symmetry property
\begin{equation}
\left( E_{\mu}^{(-)} \Vd \V \right)^{ij,kl,mn}_{-\kt} =\left( \left( E_{\mu}^{(-)} \Vd \V \right)^{ji,nm,lk}_{\kt} \right)^{*}\;.
\end{equation}
Based on this procedure, the small-x gluon TMDs in the $gg$ channel are then given by
\begin{subequations}
\begin{align}
\frac{g^2 (2\pi)^3}{4 a_s^2} F^{(1)}_{gg}(\kt) &=\frac{1}{N_c}~\left|  \left( E_{\mu}^{(-)} \Vd \V \right)^{im,mj,kl}_{\kt}  \right|^2 \;,\\
%\frac{g^2 (2\pi)^3}{4 a_s^2} F^{(2)}_{gg}(\kt) &=& \frac{1}{N_c}~ \left(\left(E_{\mu}^{(-)} \Vd \V \right)^{ij,kl,mi}_{\kt}\right) \left( \left(E_{\mu}^{(-)} \Vd \V \right)^{ki,il,mj}_{\kt} \right)^{*} \;,\\
\frac{g^2 (2\pi)^3}{4 a_s^2} F^{(2)}_{gg}(\kt) &= \frac{1}{N_c}~ \left(\left(E_{\mu}^{(-)} \Vd \V \right)^{mi,jk,lm}_{\kt}\right) \left( \left(E_{\mu}^{(-)} \Vd \V \right)^{jm,mk,li}_{\kt} \right)^{*} \;,\\
\frac{g^2 (2\pi)^3}{4 a_s^2}  F^{(3/4)}_{gg}(\kt)&= \left|  \left( E_{\mu}^{(-)} \Vd \V \right)^{lk,ki,jl}_{\kt} \right|^{2} \;,\\
\frac{g^2 (2\pi)^3}{4 a_s^2}  F^{(5)}_{gg}(\kt) &= \left(\left(E_{\mu}^{(-)} \Vd \V \right)^{ij,kl,mn}_{\kt}\right) \left( \left(E_{\mu}^{(-)} \Vd \V \right)^{kn,il,mj}_{\kt} \right)^{*}  \;,\\
\frac{g^2 (2\pi)^3}{4 a_s^2}  F^{(6)}_{gg}(\kt) &= \frac{1}{N_c^2}~\left|  \left( E_{\mu}^{(-)} \Vd \V \right)^{ij,kl,mn}_{\kt}  \right|^2\;,
\end{align}
\end{subequations}
such that all the TMDs can be calculated as local operators in Fourier ($\kt$) space.

\bibliographystyle{JHEP}
% \bibliographystyle{unsrt}
% \bibliography{refs}
\bibliography{paper}

\end{document}